%% file: main.tex
 \documentclass[manuscript, nonacm]{acmart} 

\usepackage{makecell}
\usepackage{placeins}
\usepackage{multirow}

\AtBeginDocument{%
  }

\setcopyright{acmlicensed}
\copyrightyear{2026}
\acmYear{2026}
\acmDOI{XXXXXXX.XXXXXXX}
\acmConference[IUI '26]{29th ACM Conference on Intelligent User Interfaces}{March 23--26, 2026}{Paphos, Cyprus}

\begin{document}

\title{Deliberate Lab: A Platform for Real-Time Human–AI Social Experiments}
\author{Crystal Qian}
\authornote{Both authors contributed equally to this research.}
\affiliation{%
  \institution{Google DeepMind}
  \city{New York City}
  \country{USA}
}

\author{Vivian Tsai}
\authornotemark[1]
\affiliation{%
  \institution{Google DeepMind}
  \city{New York City}
  \country{USA}
}

\author{Michael Behr}
\affiliation{%
  \institution{Google}
  \city{Cambridge}
  \country{USA}
}
\author{Nada Hussein}
\affiliation{%
  \institution{Google DeepMind}
  \city{Cambridge}
  \country{USA}
}

\author{Léo Laugier}
\affiliation{%
  \institution{EPFL}
  \city{Lausanne}
  \country{Switzerland}
}
\author{Nithum Thain}
\affiliation{%
  \institution{Google DeepMind}
  \city{Toronto}
  \country{Canada}
}

\author{Lucas Dixon}
\affiliation{%
  \institution{Google DeepMind}
  \city{Paris}
  \country{France}
}


\renewcommand{\shortauthors}{Qian et al.}
\newcommand{\todo}[1]{\textcolor{blue}{TODO: [#1]}}

\begin{abstract}
Social and behavioral scientists increasingly aim to study how humans interact, collaborate, and make decisions alongside artificial intelligence. However, the experimental infrastructure for such work remains underdeveloped: (1) few platforms support real-time, multi-party studies at scale; (2) most deployments require bespoke engineering, limiting replicability and accessibility, and (3) existing tools do not treat AI agents as first-class participants. We present \textit{Deliberate Lab}, an open-source platform for large-scale, real-time behavioral experiments that supports both human participants and large language model (LLM)-based agents. We report on a 12-month public deployment of the platform (N=88 experimenters, N=9195 experiment participants), analyzing usage patterns and workflows. Case studies and usage scenarios are aggregated from platform users, complemented by in-depth interviews with select experimenters. By lowering technical barriers and standardizing support for hybrid human-AI experimentation, \textit{Deliberate Lab} expands the methodological repertoire for studying collective decision-making and human-centered AI.\footnote{https://github.com/PAIR-code/deliberate-lab.}


\end{abstract}



\keywords{Social computing, large language models, open-source software, computational social science}

\begin{teaserfigure}
\includegraphics[width=\textwidth]{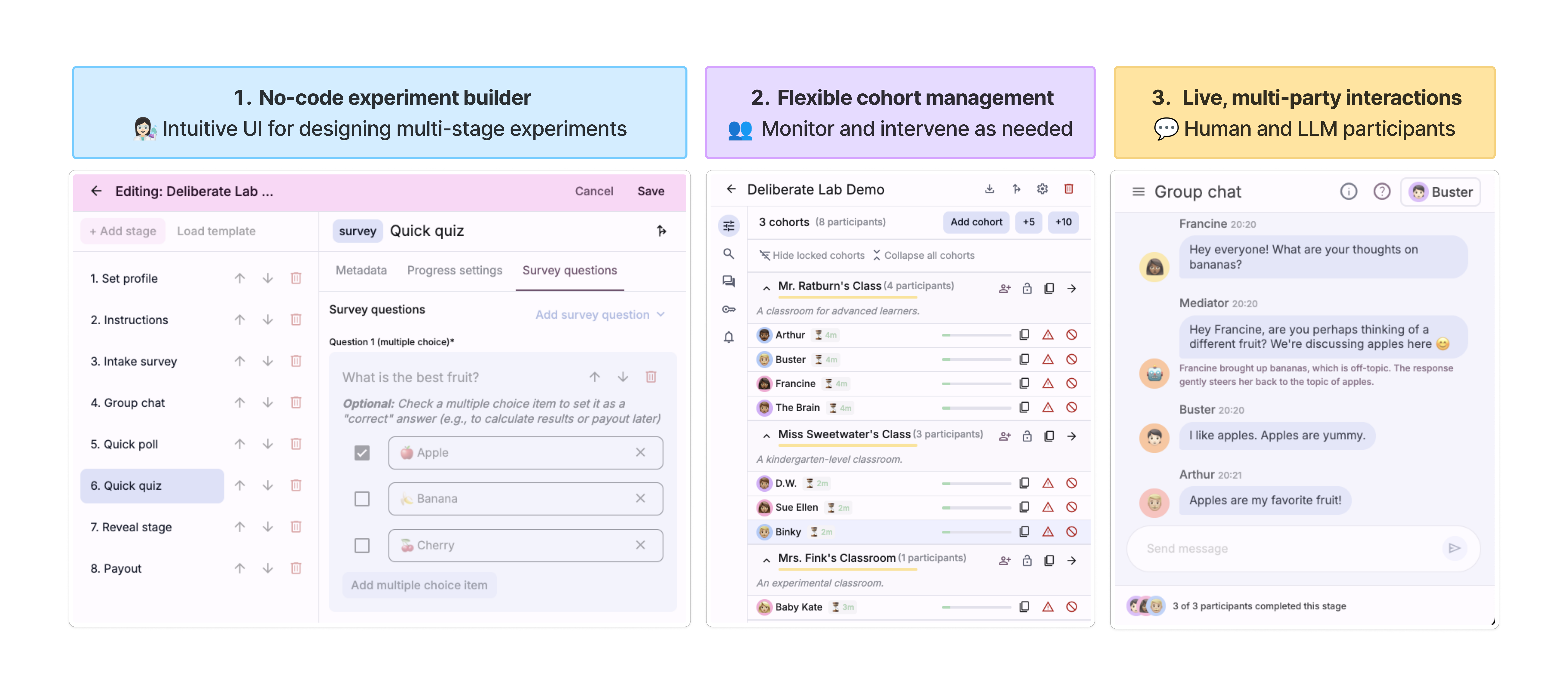}
  \caption{\textit{Deliberate Lab} system overview. Experimenters can (1) create, (2) facilitate, and intervene in real-time online experiments with human and LLM-based participants. (3) shows a live conversation interface involving human participants and an LLM mediator.}
  \Description{}
  \label{fig:teaser}
\end{teaserfigure}


\maketitle

\input{sections/01_intro}

\input{sections/02_related_work}

\input{sections/03_cuj}

\input{sections/05_system_overview}

\input{sections/07_usage}
\input{sections/08_case_studies}
\input{sections/09_limitations}

\input{sections/10_conclusion}

\clearpage

\begin{acks}
We'd like to thank James Wexler, Michael Xieyang Liu, Fernanda Viegas, Rasmi Elasmar, Ian Baker, Thibaut De Saivre, the People and AI Research team, our interview participants, and the Deliberate Lab community. Thank you for your feedback and contributions.

\end{acks}


\bibliographystyle{ACM-Reference-Format}
\bibliography{main}

\newpage

\appendix

\input{sections/appendix}

\end{document}

%% file: sections/01_intro.tex
\section{Introduction}\label{sec:intro}

The structure of social interaction is increasingly digital. Collective behaviors such as deliberation, negotiation, and coordination increasingly take place within online systems that are shaped by algorithmic and interface constraints. This calls for expanded empirical study of how groups interact, reason, and converge in digitally-mediated environments \citep{costello2024, tessler2024}, yet current methodological tools remain poorly aligned with these settings, offering limited support for real-time, multi-participant experimentation.

Existing platforms for behavioral research are optimized for single-user, asynchronous paradigms, such as surveys \citep{gorilla, qualtrics}. They often require bespoke engineering and platform development for customization \citep{Almaatouq2021}, which can hinder rapid and reproducible experimentation, particularly for researchers without extensive technical resources. Furthermore, as interest grows in hybrid human-agent systems and algorithmically mediated group interaction \citep{horton2023, park2022social}, there is a need for systems to consider digital or synthetic participants as well as traditional humans \citep{prolific, concordia2023, park2023generative}.


We introduce \textbf{Deliberate Lab}, a no-code, open-source platform for real-time social experimentation. Designed to support synchronous interaction, with large language models as first-class participants, Deliberate Lab lowers the technical barrier for interdisciplinary research while enabling reproducibility, extensibility, and flexible experimental design. The platform accommodates a wide range of configurations, from purely human deliberation to hybrid collectives involving LLM agents as participants, moderators, or simulated crowds.

During a 12-month public access period, researchers with diverse disciplinary backgrounds and varying levels of technical expertise used the platform to support a wide range of experiments and user studies. Psychologists conducted large-scale online elections, economists modeled negotiation dynamics with AI interventions, and HCI researchers extracted structured data on human–AI collaboration. Additional use cases included developing AI teachers and debate moderators, conducting remote product focus groups, and exploring real-time adjudication for data annotation. These applications are described in detail and supplemented with insights from user interviews.

The primary contributions of this paper are as follows:

\begin{enumerate}
\item \textbf{System design and architecture.} Novel architectural choices that operationalize real-time, synchronous human–AI interaction, including rich facilitation features like cohort transfers and real-time waiting stages, and multi-agent conversational considerations like artificial throttling (``WPM'') and hand-raising.

\item \textbf{Deployment analytics.} Quantitative findings and usage patterns from a 12-month open-access deployment period, involving 88 experimenters, 597 created experiments, and 9,195 participants.

\item \textbf{Case studies and user insights.} Qualitative findings from experimenter feedback and interviews, highlighting cross-disciplinary adaptations, usability challenges, and emerging research opportunities.

\end{enumerate}

Taken together, these contributions demonstrate how Deliberate Lab contributes towards making real-time, hybrid human–AI experimentation a standard tool in the social and behavioral sciences.

%% file: sections/02_related_work.tex
\section{Related Work}\label{sec:related_work}

\subsection{Experiments in collective behavior}

The study of collective human behavior builds on a long tradition of experimental paradigms designed to isolate specific social dynamics. Classic examples include strategic games such as the Prisoner’s Dilemma, which tests bilateral cooperation~\citep{axelrod1981evolution}, public goods games, which examine collective action~\citep{fehr2000cooperation}, and foundational group studies evaluating polarization and psychological safety~\citep{moscovici1969group, edmondson1999psychological}. As social behaviors have moved online, research has examined how digital environments and computational methods can shape, analyze, and model group dynamics~\citep{dowell2019group, soliman2024}.

Advances in large language models (LLMs) have further expanded this space, introducing new patterns of hybrid human–AI interaction~\citep{lee2022evaluating, 10.1145/3640543.3645198}. LLM-based interventions have shown potential in improving individual and group processes, such as debunking conspiracy theories~\citep{costello2024}, improving deliberation~\citep{small2023opportunitiesrisksllmsscalable, tessler2024}, and mediating political debates~\citep{argyle2023}.

Despite this progress, studies of collective behavior often rely on bespoke experimental setups, limiting reproducibility and cross-study comparison~\citep{openscience2015estimating, reproducibility2015, camerer2018evaluating, camerer2016evaluating}. Experiments involving LLMs introduce further technical overhead, typically requiring custom system architectures for model hosting and deployment. Addressing these limitations calls for shared, extensible platforms capable of supporting complex, real-time, and hybrid experimental designs.

\subsection{Platforms for behavioral and social experiments}
Traditional platforms for online behavioral research, such as Qualtrics, Gorilla, or recruitment platforms like Prolific and Amazon Mechanical Turk, are primarily designed for asynchronous, single-participant tasks \citep{qualtrics, gorilla, prolific, mechanicalturk}. While powerful for surveys and individual cognitive studies, they lack the infrastructure to support real-time, group-based interaction. Specialized platforms have been developed to address this gap. Empirica \citep{Almaatouq2021} provides a key open-source framework for developing and running real-time, multi-user experiments. Tools like zTree \citep{fischbacher2007z} and oTree \citep{chen2016otree} are widely used in experimental economics and psychology for interactive strategic games. While these platforms are foundational for multi-party human research, they require substantial programming expertise to configure. Furthermore, they were not designed with first-class support for LLM-based agents or facilitators --- an emerging area of research that bridges human and computational social experimentation.

\subsection{Simulating social dynamics with LLM agents}
Parallel to platforms for human experiments, a growing line of research uses LLMs to create ``in-silico'' simulations of social environments, supported by generative modeling libraries such as Concordia~\citep{concordia2023} and EDSL~\citep{edsl2024}. A prominent example is the work of \citet{park2023generative}, who developed a sandbox environment populated by generative agents exhibiting believable individual and social behaviors. Other studies have simulated social media discussions~\citep{park2022social}, generated task-oriented dialogues through dyadic role-play~\citep{camel2023}, explored how agent personalities influence debate outcomes~\citep{taubenfeld2024}, and demonstrated human-like behavior in economic games~\citep{horton2023}. While these multi-agent simulations provide valuable testbeds for studying social dynamics, they remain detached from human behavior, motivating the need for complementary work on real human–AI interaction.

\subsection{Frameworks for human-AI group interaction}
Emerging systems enable humans to interact directly with groups of AI agents, bridging the gap between human-only experiments and fully autonomous simulations. For example, \citet{gu2024agentgroupchatinteractivegroupchat} introduced a platform where a single human engages with multiple agents in structured tasks such as debates and inheritance disputes, while SAUCE~\citep{neuberger2024saucesynchronousasynchronoususercustomizable} allows a user to interact with a customizable group of LLMs. While these tools demonstrate the feasibility of human-in-the-loop interaction with agent groups, they are typically limited to single-participant settings and rigid task structures, constraining their use for the flexible, multi-human, multi-agent designs common in behavioral science. 

Deliberate Lab addresses this remaining gap. It provides a configurable, open-source platform that supports real-time, synchronous interactions among \textit{any number} of human and LLM participants. By enabling flexible experimental design, the platform empowers researchers of any technical background to investigate human–AI group dynamics at a scale and fidelity that is not easily achievable with existing tools.

%% file: sections/03_cuj.tex
\section{User Journeys}\label{section:user_journeys}

We outline three core user journeys in Deliberate Lab: (1) configuring experiments, (2) facilitating experiments in real time, and (3) participating in experiments. Table \ref{tab:core-concepts} provides key concepts for understanding these interactions.

\begin{table}[h]
\renewcommand{\arraystretch}{1.5}
\centering
\small
\begin{tabular}{p{0.2\linewidth} p{0.7\linewidth}}
\toprule
\textbf{Term} & \textbf{Definition} \\
\midrule
\textbf{Experiment} & A synchronous interaction sequence with configurable components, metadata, and optional LLM integrations. \\

\textbf{Template} & A reusable experiment configuration designed to accelerate experiment setup. \\

\textbf{Stage} & A module within an experiment (e.g., survey stage, election stage) with configurable logic and interface. Appendix~\ref{app:architecture-stages} includes a list of supported stages. \\

\textbf{Participant} & A human or LLM agent that progresses through an experiment. \\

\textbf{Cohort} & A group of participants that progress through an experiment together, sharing access to data in group-oriented stages (e.g., group chat, elections). Each participant is assigned to a single cohort and can be dynamically transferred between cohorts mid-experiment. Multiple cohorts may run in parallel. \\

\textbf{Mediator} & An LLM agent mapped to a single stage (e.g., in a group chat stage or private 1:1 interaction). Mediators can be granted access to data (including participant answers) from other stages. \\

\textbf{Prompt} & A structured instruction provided to an LLM agent participant or mediator, assembled from global and stage-specific components. \\
\bottomrule
\end{tabular}
\caption{Core concepts and terminology used throughout the Deliberate Lab platform.}
\label{tab:core-concepts}
\end{table}

\begin{figure}[ht]
  \centering
  \includegraphics[width=\linewidth]{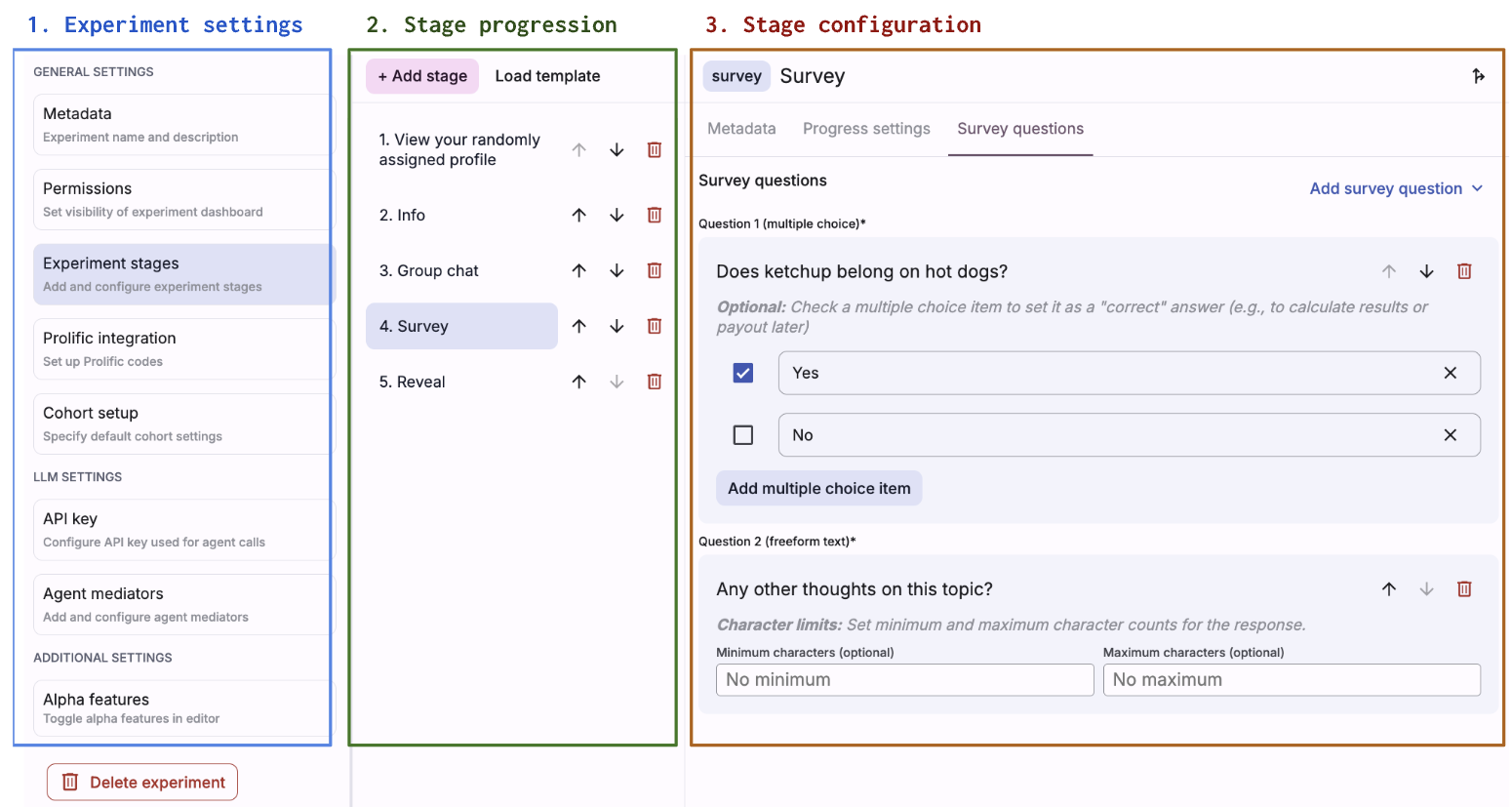}
  \caption{Experiment creation view. (1) Settings configuration, such as metadata and LLMs. (2) Stage ordering panel; stages can be added manually (full list in Appendix~\ref{app:architecture-stages}) or loaded from a template. (3) Stage editing panel; a survey question editor is shown.}
  \Description{Screenshots of the Deliberate Lab experiment creation view, showing experiment settings, stage ordering panel, and stage editing panel.}
  \label{fig:creation_view}
\end{figure}

\subsection{Configuring an experiment}

Experimenters can create a new experiment in three ways: (1) by forking an existing experiment; (2) by selecting a template from the homepage gallery, which includes common configurations and community-contributed designs (Figure~\ref{fig:home_view}); or (3) by building one from scratch using modular stages (Figure~\ref{fig:creation_view}, Figure~\ref{fig:creation_view2}). Each stage is configurable with markdown-rendered content, optional visual elements (e.g., descriptions, progress bars), and interaction parameters such as timers for auto-advancing or synchronized waiting screens for group stages (e.g. chats, elections).

Experiments can be designated as ``public,'' enabling any authenticated experimenter with the link to view, assist in facilitation, or fork the configuration without editing privileges. Integration with the Prolific recruitment platform enables the use of redirect and completion codes to automate participant onboarding and compensation \citep{prolific}.

\subsubsection{Adding experiment stages.} A full list of out-of-the-box experiment stages is provided in Appendix~\ref{tab:experiment-stages}. Here, we describe a few stages that enable core user journeys or highlight novel design considerations.

\begin{itemize}
    
    \item \textbf{Chat.} Chat stages enable real-time messaging between participants and agents. High participant engagement in chat stages during initial pilot experiments prompted the introduction of timers to manage session length. The LLM implementation for real-time messaging is detailed in Section~\ref{sec:llms}.
    
    \item \textbf{Survey.} Survey stages include freeform questions, sliders, multiple choice questions, and quizzes. \textit{Per-participant survey stages} were additionally requested by the experimenter community and introduced to enable dynamic question rendering based on cohort composition (e.g. questions related to each participant). \textit{Comprehension check} survey stages can be used to ensure participants understand key instructions or rules before proceeding.
    
    \item \textbf{Transfer.} Transfer stages are used to funnel all participants through a shared entry point and then sort them into smaller cohorts based on time of entry or survey responses (e.g., to ensure viewpoint diversity). This setup is commonly paired with a timeout condition to remove unmatched participants after a set period.    
    
    \item \textbf{Profile.} Visible identity cues strongly influence group dynamics. Experimenters can either (a) allow participants to choose their own name, avatar, and pronouns, or (b) assign pseudonymous profiles~\citep{soliman2024, peddinti2014}. Deliberate Lab offers several profile sets, including animal-themed (e.g., Anonymous Owl), nature-themed (e.g., Anonymous Mountain), and numeric (e.g., Anonymous 5192). A study leveraging these identifiers is described in Section~\ref{sec:lost_at_sea}.
    
    \item \textbf{Reveal.} Reveal stages are used to debrief participants or present outcomes; they can display individual responses, cohort-level summaries, or collective outcomes (e.g., election winners, group survey responses).
    
    \item \textbf{Payout.} Designed with online recruitment platforms in mind, payout stages enable highly flexible incentive structures. Rewards can be tied to stage completion, quiz performance, or randomized conditions. Participant payouts can be exported in CSV format to streamline compensation workflows via tools like Prolific.

\end{itemize}

\subsubsection{Adding LLM agents.} Agents can be added either as synthetic \textit{participants} or as within-stage \textit{mediators}. Mediator agents are defined through a set of prompts that provide stage-specific instructions, optionally referencing context from prior stages. This allows for roles like a ``Friendly Facilitator'' or a ``Private Expert Coach'' who adapts to prior chat history. Participant agents use the same underlying scaffolding, but also receive a persona prompt when added to an experiment (e.g. ``Bob the Friendly Facilitator.’’).

\begin{figure}[ht]
  \centering
  \includegraphics[width=0.5\linewidth]{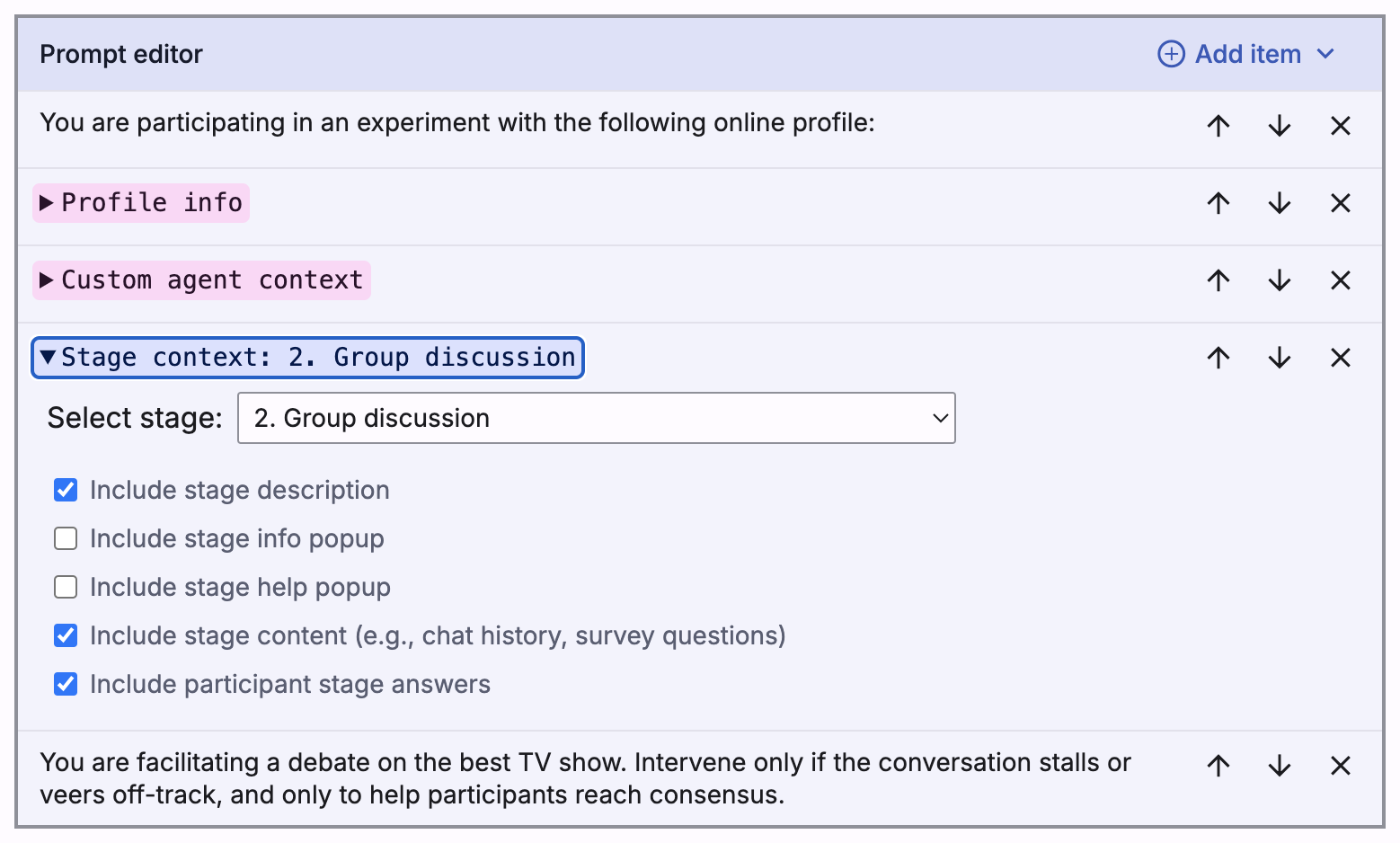}
  \includegraphics[width=0.45\linewidth]{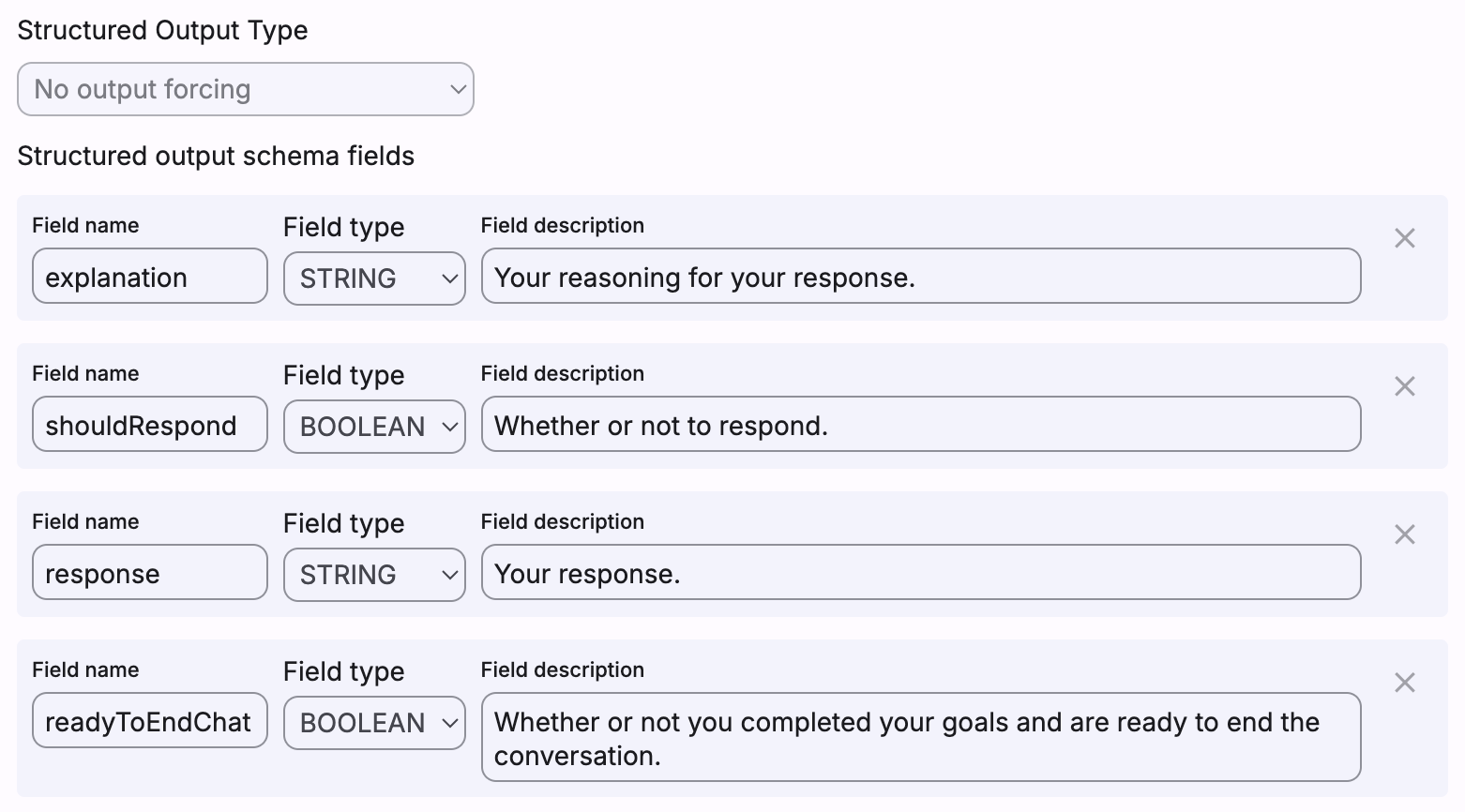}
  \caption{Prompt editor for chat stage. On the left, a list editor with prompt items enables the experimenter to define and reorder custom text as well as variables like stage context. On the right, the experimenter can use the dropdown to select a structured output mode and/or customize structured output fields.}
  \Description{Screenshots of the prompt editor for chat stage. In the left screenshot, a list editor with prompt items enables the experimenter to define and reorder custom text as well as variables like stage context. In the right screenshot, there is a dropdown for selecting a structured output mode and a list of editable structured output fields.}
    \label{fig:prompt-editor}
\end{figure}

Agents can work out of the box with minimal configuration; for example, a mediator can be added to a chat stage with a simple prompt such as ``ensure politeness.'' However, agent configuration can be richly customized through modular prompt components, including modifying the agent’s profile, optional custom instructions, and selected context from previous stages (Figure~\ref{fig:prompt-editor}). These components can be added, reordered, or removed to tailor the prompt logic. Figure~\ref{fig:prompt-template} shows an example of the prompt scaffolding for an in-chat mediator.

\paragraph{Supported LLMs.} Deliberate Lab supports API keys for any model compatible with the Chat Completions API format, with explicit UI support for Google Gemini and OpenAI GPT. This format integrates with gateways such as OpenRouter and LiteLLM, enabling access to a broad range of models. A connector for custom-hosted models on Ollama was contributed by the open-source community~\citep{ollama2024}.

\begin{figure}[ht]
  \centering
  \includegraphics[width=\linewidth]{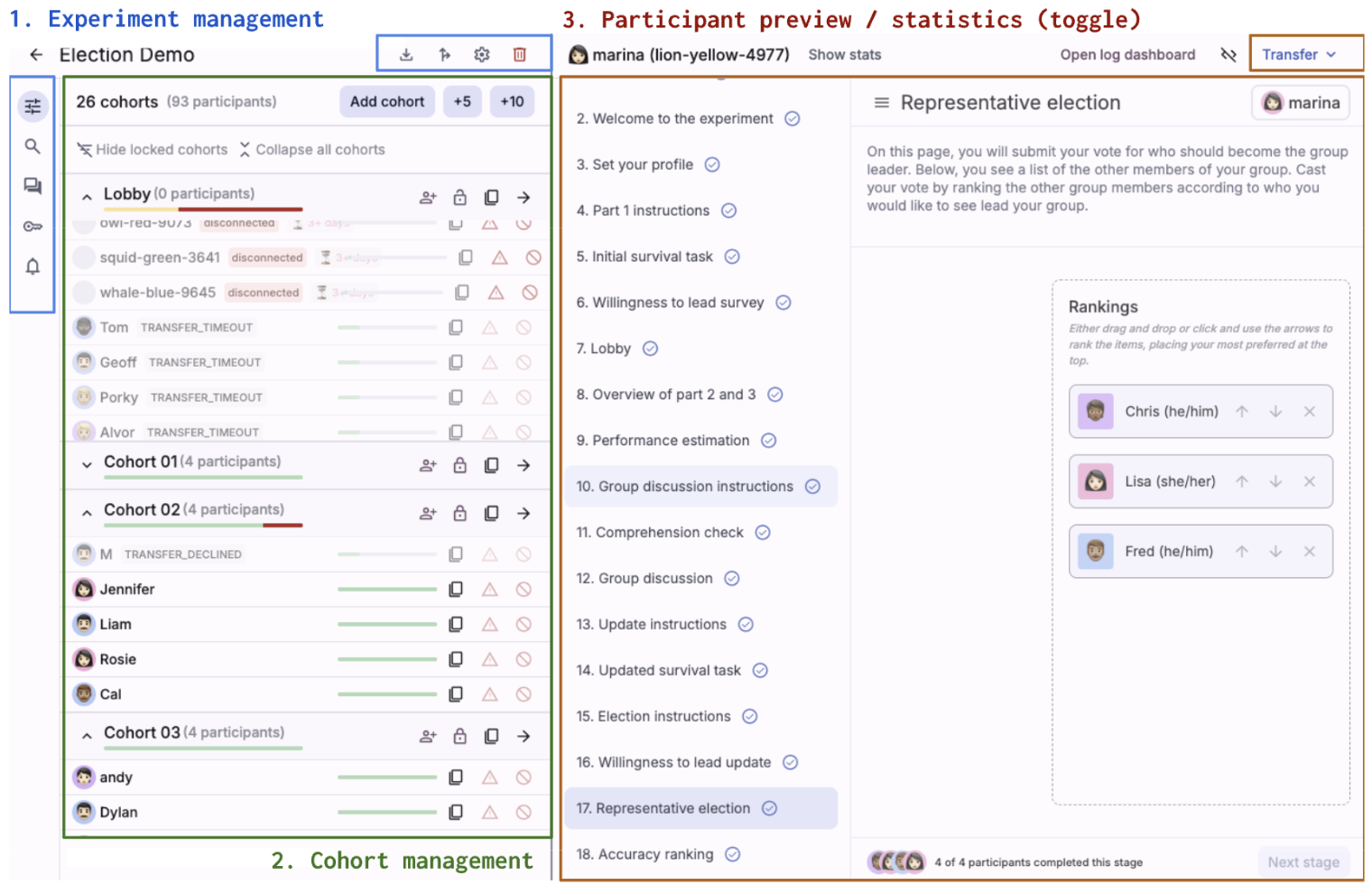}
  \caption{Facilitator view for a ``lobby''-style experiment. (1) Experiment management toggle, including a search bar to find participants, a messaging feature to message participants, LLM API configuration settings, and a notification panel to address active assistance requests. (2) On the left, a cohort management panel showing all participant groups, with live participant monitoring and attention check features. (3) On the right, an experiment preview panel shows a selected participant's  live view; experimenters can view and submit input on the participant's behalf if needed, and toggle debugging messages for LLMs (visible only to the experimenter).}
  \label{fig:facilitator_view}
  \Description{Screenshots of the facilitator view for a Deliberate Lab experiment. The left panel displays experiment management features and a cohort management panel showing all participant groups. The right panel displays a live interface for a selected participant, allowing experimenters to view participant statistics and inputs and submit input on a participant's behalf.}
\end{figure}

\subsection{Facilitating an experiment}

Figure~\ref{fig:facilitator_view} shows the experiment dashboard, which enables experimenters to preview and manage participants in real time. Experiment metadata (Prolific integration, description, etc.) can always be edited; the stage configuration can be edited until a participant joins.

\paragraph{Adding participants.} Participants can be added to experiment \textit{cohorts}, groups that progress through the experiment together with shared group data. Human participants join through shared URLs, without login requirements. Agent participants can be added to a cohort by (1) selecting an agent template constructed during the experiment creation process (e.g., "Friendly Facilitator"), and (2) adding optional participant-level customization and backstory.

\paragraph{Transferring cohorts.}\label{sec:transferring}
Experimenters can manually transfer participants between cohorts or remove them entirely during a session, enabling dynamic reconfiguration. This feature is frequently used in ``lobby'' style experiments, where participants first join an initial lobby cohort to complete asynchronous onboarding tasks such as reading instructions or passing comprehension checks. Participants may then be transferred to new cohorts for several reasons: to balance group composition (e.g., matching participants with opposing views or complementary attributes), to support rotating or multi-round designs where participants cycle through different groups across stages, or to match participants who are active simultaneously for synchronous interaction.
 
\paragraph{Real-time monitoring.} Maintaining participant engagement is critical in online, collaborative experiments, as a single dropout can disrupt or invalidate an entire group’s data. This challenge is amplified in online settings, where attention and retention are harder to control. To address this, real-time monitoring tools were developed in collaboration with experimenters conducting live studies and refined through their feedback. \textit{Status indicators} flag when a participant is inactive or lagging behind, allowing experimenters to trigger attention checks. \textit{Presence detection} displays whether each participant is connected, disconnected, or active, helping facilitators decide when to intervene, transfer, or remove users. If a participant remains unresponsive, they can be removed and replaced mid-session using the ``boot'' control, preventing dropouts from disrupting group progress.

\paragraph{Responding to participant needs} In chat-based stages, facilitators can message participants directly to resolve confusion or notify them of technical issues. Additional features were developed to support large-scale experiments, where a single experimenter cannot monitor every group in real time. Participants can trigger an \textit{alert}, which sends a notification to the facilitator for timely assistance. A \textit{participant search} function allows experimenters to quickly locate individuals by name, pseudonymous profile, or external ID (e.g., Prolific ID), enabling responsive support even when participants reach out through external channels.

\paragraph{Monitoring agents.} Experimenters can monitor LLM agents via an agent debugging panel (Figure~\ref{fig:log_debugging}), which surfaces LLM call logs and allows agents to be paused mid-experiment. Agent settings like prompts and API selections can be adjusted in real time, which is useful for testing or debugging agent configurations and behaviors.

\paragraph{Post-experiment facilitation} Once cohorts or experiments reach capacity, facilitators can lock them to prevent new participants from joining. A .zip archive can be downloaded at any time, containing the experiment configuration, structured JSON logs, and CSV exports of stage data including chat transcripts, survey responses, and participant metadata. For incentivized experiments with Prolific integration, a formatted CSV of participant IDs mapped to completion status and bonus payouts enable quick payment processing.

\subsection{Participating in an experiment}

The participant view (Figure~\ref{fig:facilitator_view}, Figure~\ref{fig:participant_view}) includes a collapsible stage navigation menu (left), a main content area (right), and a profile display with help and alerting icons. Participants advance through stages by clicking “Next stage” and can revisit previous stages at any time. When enabled, wait screens delay content for synchronized stages until all group members have arrived, while cohort progress bars visualize peers’ real-time progress. Participants can also accept or decline transfers, respond to attention checks, or trigger an alert to notify the experimenter of issues. After completing the final stage, they are routed to an end screen or a pre-configured URL, such as a Prolific completion link.





%% file: sections/05_system_overview.tex
\section{System Design}\label{sec:architecture}

\subsection{System architecture}

Deliberate Lab is a TypeScript-based web application with a modular frontend and backend. Both layers share internal packages for constants, data structures, and utility functions.

\begin{figure}[ht]
  \centering
  \includegraphics[width=.4\linewidth]{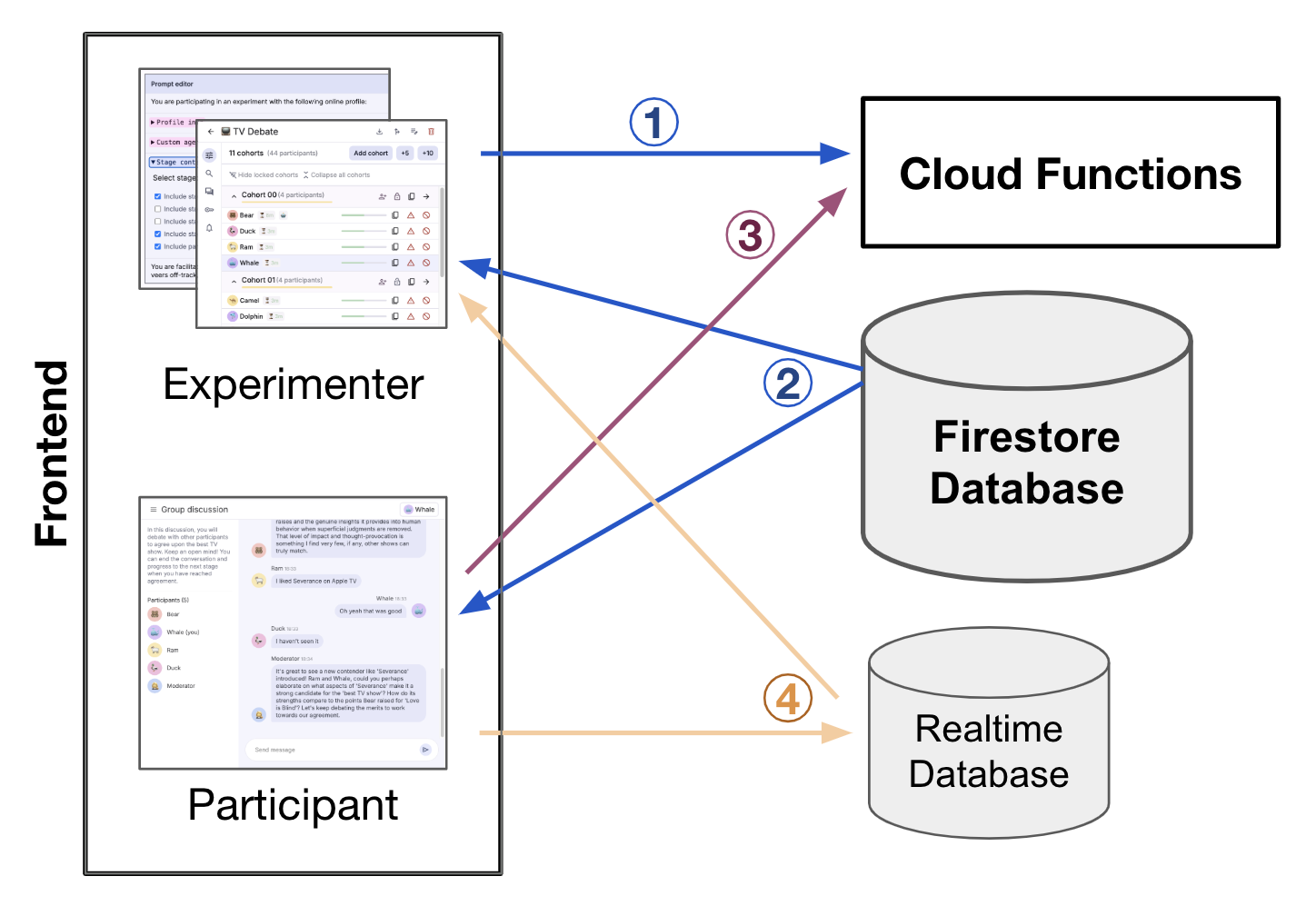}
  \caption{
    Simplified architecture of the Deliberate Lab platform stack. (1) Experiments are configured and submitted to a backend endpoint, (2) then stored in the primary Firestore database (Figure~\ref{fig:system-architecture}) for frontend access. (3) Participant inputs are sent to the backend and logged in the primary Firestore database. (4) Presence signals are written to the secondary Realtime database to track active users.
  }
  \label{fig:simple_infrastructure}
  \Description{Diagram of the Deliberate Lab stack showing experiment submission, data storage, and real-time presence tracking.}
\end{figure}

\subsubsection{Frontend}

The UI is built using the Lit Element library \citep{lit_web}. This includes larger views such as the experiment editor and experimenter dashboard, as well as stage-specific components; each stage typically has an ``editor'' component for the experiment-building user journey and a ``view'' component for the participant-navigation user journey.

MobX handles state management across both experimenter and participant flows \citep{mobx_readme}. The system includes an service for authentication, backend-connected services for logged-in experimenters to configure and manage experiments, and a set of participant-facing services for reading the current experiment, public cohort data (including pseudonymous profiles), and private participant state. A separate ``participant answer service'' caches client-side inputs locally until they are explicitly submitted to the backend via manual click events.

\subsubsection{Backend}
The backend was built with Google Firebase platform \citep{firebase_firestore_docs} and includes two databases and a set of Cloud Functions endpoints.

\begin{enumerate}
\item \textbf{Primary Firebase database.} Stores all experiment-related content, including allowlist email addresses for experimenter log-in, experiment configurations, and participant-generated data (e.g., survey responses, chat messages). Data is written via Cloud Functions, both directly through endpoints and indirectly via endpoint triggers.
\item \textbf{Secondary Realtime database.} Tracks participant presence (i.e., whether a participant is actively viewing a Deliberate Lab page). Experimenters can leverage this information, visible in the experiment dashboard, to handle inactive users (e.g. send attention checks, deprioritize for cohort transfer).
\item \textbf{Cloud Functions.} Callable endpoints are invoked by the frontend when experimenters update configurations or when participants submit input. These trigger functions handle:
    \begin{enumerate}
    \item \textbf{Cohort state.} When a participant submits private data (e.g. a vote in an election stage), a trigger updates any relevant public cohort-level data (e.g., recalculation of the election winner) in the primary Firebase database.
    \item \textbf{Agent logic.} When a new chat message is written, a trigger prompts relevant agents to optionally respond. When an agent participant is created or reaches a new stage, a trigger calls the corresponding agent logic to complete the stage and move forward.
    \end{enumerate}
\end{enumerate}


\subsubsection{Security and privacy}
Security and privacy were central design considerations, especially given the intent to maintain a publicly accessible instance of Deliberate Lab. We consulted with domain experts and completed a Privacy and Data Protection (PDD) review within our institution to ensure that the system aligns with standard ethical protocols for online behavioral research.

\paragraph{Experimenter permissions.} Experimenters gain access when logged in with an email address that is included in the Firestore's manually-configured allowlist. All others are treated as participants. Any experiment marked “public” can be shared with other verified experimenters. Experimenters with manually-configured administrator access in Firestore have full access to view and delete all experiments via an administrator-only dashboard.

\paragraph{Experiment management.} All experimenter actions, such as saving configurations or managing cohorts, are secured via authentication checks in backend functions. Firestore rules enforce that for each Deliberate Lab experiment, only experimenters with explicit roles (creator, editor, or reader) can read experimenter-specific data, and only the creator or administrators can write to or delete the experiment configuration.

\paragraph{Experiment participation.} Participants do not need to log in. Each participation link contains a UUID-based private ID used to access that session, while all shared references (e.g., chat displays) rely on a separate public ID. This separation makes it unlikely for anyone besides the experimenter and participant to access sensitive identifiers.
    
\paragraph{LLM APIs.} Experimenters' LLM API keys are stored in secure, authenticated, experimenter-level configurations. When an agent stage triggers a model call, the backend fetches the relevant API key based on the experiment creator’s credentials, ensuring that keys are only used by their owners and not exposed to other users or participants.
    
\subsection{Real-time, synchronous LLM integration}\label{sec:llms}

Enabling real-time human–LLM interaction was a core challenge in Deliberate Lab’s design. This section details the implementation choices that enabled synchronous, hybrid experiments.

\paragraph{Context and memory management.}
Early platform implementations were constrained by limited context windows, requiring truncation of chat history in agent prompts. With advancing model capacities, full conversation transcripts with scaffolding easily fit within context windows. Future work may explore structured memory systems to improve efficiency and contextual grounding \citep{park2023generative}.

\paragraph{Prompt design for structured agent behavior.}
Several agent behavior issues required explicit prompt adjustments:
\begin{itemize}
    \item \textbf{Message ordering.} Initial agent responses often failed to account for message ordering in the provided chat transcripts, reacting disproportionately to older messages or ignoring recent context. To improve coherence, chat history is formatted with speaker, timestamp, and message fields, and prompt instructions explicitly direct agents to reason sequentially over the dialogue.

    \item \textbf{Selective intervention.} Agents initially responded to nearly every message, reducing the effectiveness of interventions. Structured output fields like \texttt{shouldRespond} and \texttt{readyToEndChat} were introduced to enable more targeted behavior. These fields are customizable; for instance, a safety agent might include a \texttt{severityScore} (e.g., 1–5) structured output parameter and only respond if the score exceeds a threshold.
    
    \item\textbf{Pseudonym-induced bias.} Pseudonymous profiles (e.g., “Anonymous Bear”) occasionally triggered thematically biased or playful responses, such as puns or emojis. Default prompts now include explicit instructions to ignore name-related semantics in pseudonymous avatar sets.
\end{itemize}

\paragraph{Human-centered interaction design.} 
To support natural, multi-turn chat with real-time human participants, several design adjustments were made. Unlike traditional multi-agent-only systems \citep{concordia2023}, which use fixed turn-taking, Deliberate Lab initially implemented a "hand-raising" system: if multiple agents wanted to respond, one was randomly selected while the others' responses were logged for debugging. However, the introduction of faster models led to agents responding too quickly and dominating the conversation. To mitigate this, a configurable “words per minute” (WPM) setting was added to simulate human-like typing delays. Based on experimenter feedback, WPM was later incorporated into the response prioritization algorithm, enabling faster agents to respond more often. This hybrid system maintains realism and can be disabled entirely in all-agent environments.

%% file: sections/07_usage.tex
\section{Field deployment and usage}\label{sec:usage}

In October 2024, a public instance of Deliberate Lab was deployed on Google Cloud Platform and announced through academic and industry mailing lists across social psychology, HCI, and related fields. Open access was granted to all approved applicants. This section reports aggregate, non-identifiable usage data from the 12-month deployment period.

\begin{figure}[ht]
  \centering
  \includegraphics[width=\linewidth]{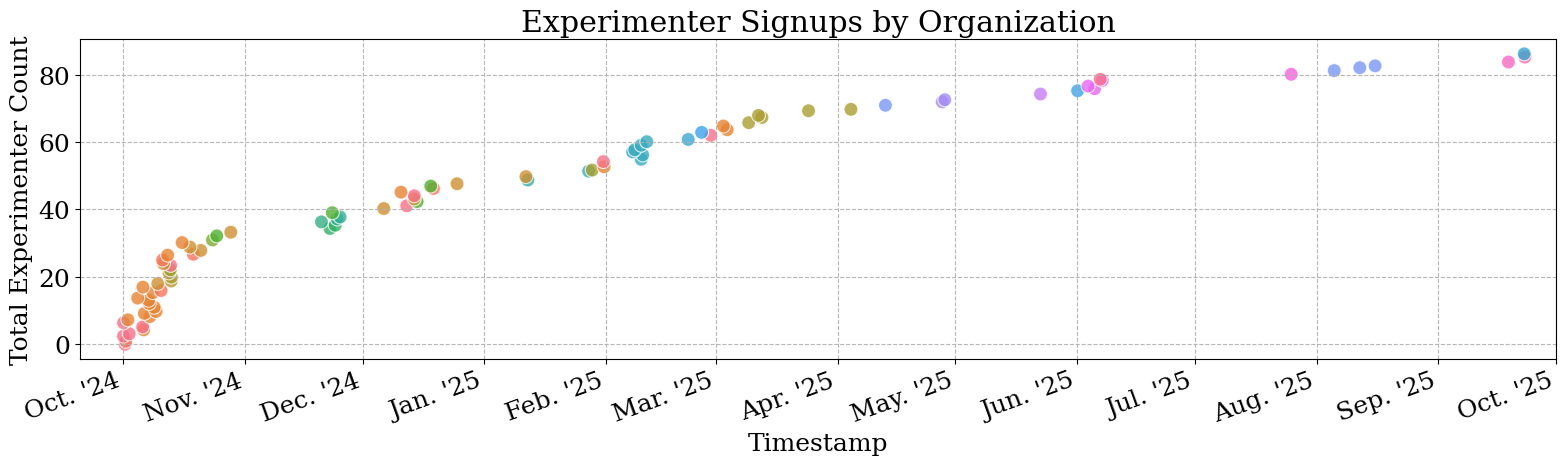}
  \caption{Cumulative experimenter signups over a 12-month period, color-coded by organization; there were 88 account applications across 26 organizations. The bursty pattern reflects waves of coordinated institutional uptake.}
  \label{fig:experimenter_signups}
  \Description{A scatter plot displaying cumulative experimenter signups by organization over a 12 month period, color-coded by affiliation / organization. Adoption was bursty but consistent, often occurring in institutional clusters.}
\end{figure}

\paragraph{Experimenter access and discovery} 88 experimenter accounts across 24 organizations—including universities, companies, nonprofits, and independent labs—applied to use the platform. The number of platform users is likely higher, as several accounts appear to represent entire teams (e.g., “team@” or “lab@”). In the signup form (Appendix~\ref{app:signup}), most users reported hearing about the platform via word of mouth, followed by the launch announcement and the open-source codebase. Figure~\ref{fig:experimenter_signups} shows cumulative experimenter signups over time, color-coded by organization. 

\begin{figure}[ht]
  \centering
  \includegraphics[width=\linewidth]{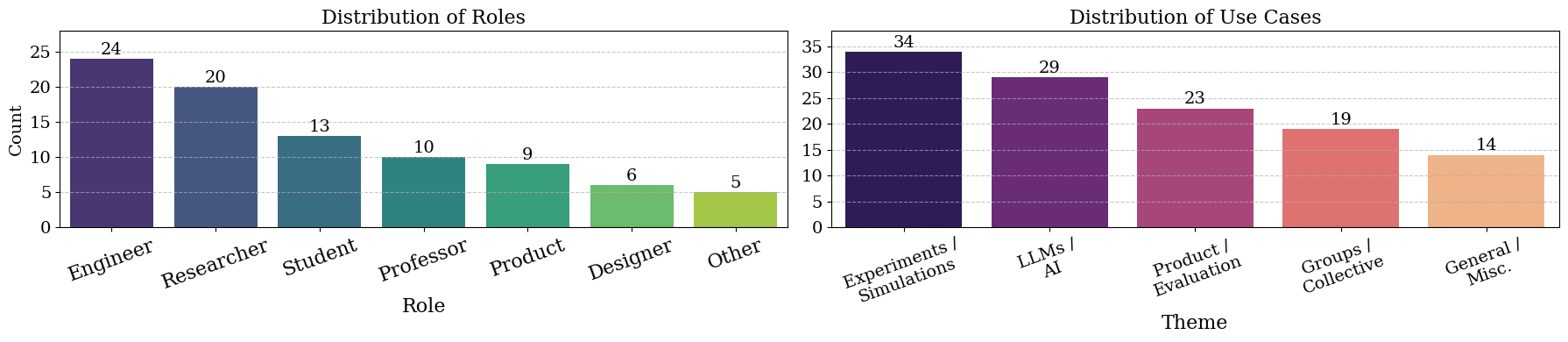}
  \caption{Experimenter characteristics. \textit{Left:} Self-reported experimenter job roles (left) and intended use cases (right).}
  \label{fig:user_distribution}
  \Description{Bar charts displaying experimenter characteristics. On the left, a bar chart displaying distribution of experimenter roles. Engineers are the most common role, followed by researchers, students, professors, product, designers, and other. On the right, a bar chart displaying distribution of use cases. Experiments/simulations are the most common use case, followed by LLMs/AI, product/evaluation, groups/collective, and general/misc. }
\end{figure}

\paragraph{Experimenter backgrounds and use cases}
The signup form collected information on experimenter backgrounds and intended use cases (Figure~\ref{fig:user_distribution}). Respondents came from a range of disciplines—including machine learning, psychology, economics, education, and policy—and spanned both technical and non-technical roles . Thematic analysis of free-text responses indicated that most experimenters planned to run behavioral studies, aligning with the platform's original framing. Representative use cases included plans to ``run experiments,'' ``study group interactions,'' or ``evaluate moderation strategies.'' 

Anticipated use cases for the LLM integrations included AI acting as a facilitator in mediation, consensus-building, or educational contexts (e.g., “test an AI tutor”) and as a simulated participant (e.g., “chat with different agent personas,” “simulate learning environments,” or “run experiments where one group member is an LLM”). Some experimenters also proposed dynamic applications, such as testing real-time generation of AI-authored statements.

Interestingly, some experimenters were drawn to the platform not for its AI integrations or social science capabilities, but simply for its multi-user interface. Unexpected use cases included product development and evaluation (e.g., assessing LLM safety or multi-agent coordination), expert annotation tasks requiring group consensus, and breakout group discussions in classroom settings. Others described multi-user data collection scenarios such as capturing public sentiment on topical issues or running structured focus groups.

\begin{table}[h!]
\centering

\label{tab:platform_stats_periods}
\footnotesize
\renewcommand{\arraystretch}{0.9}
\begin{tabular}{@{}llrllr@{}}
\toprule
\multicolumn{3}{c}{\textbf{12-month period}} & \multicolumn{3}{c}{\textbf{6-month period}} \\
\cmidrule(lr){1-3} \cmidrule(lr){4-6}
\textbf{Category} & \textbf{Metric} & \textbf{Count} & \textbf{Category} & \textbf{Metric} & \textbf{Count} \\
\midrule
\textbf{Overall} 
  & Experiments created & 597 
  & \textbf{Overall} & Experiments created & 222 \\
  & Participant entries & 9,195 
  &  & Trial experiments & 165 \\[4pt]

\textbf{Experiment creation} 
  & Templates loaded & 78 
  & \textbf{Experiment creation} & Used attention checks / timers & 21 \\
  & Forked from existing experiments & 160 
  &  & Added LLM participant templates& 28 \\
  & Created organically & 359 
  &  & Added LLM mediator templates & 60 \\[4pt]

\textbf{Cohort management} 
  & Cohorts created & 2,113 
  & \textbf{Experimenter medadata} & Unique experimenters & 38 \\
  & Transfers initiated & 2,191 
  &  & Experimenters with LLM API keys & 27 \\[4pt]
\textbf{Real-time monitoring} 
  & Participants booted & 393 
  & \textbf{Real-time monitoring} & Used Prolific integration & 81 \\
  & Attention checks sent & 527 
  &  &  &  \\
  & Attention checks passed & 395 
  &  &  &  \\
\bottomrule
\end{tabular}
\caption{Aggregate platform activity over the 12-month and 6-month periods. Twelve-month counts reflect all recorded events stored in Firebase analytics. Six-month counts represent more intentional experimental properties, measured only for experiments whose data remain under GDPR retention (e.g., inclusion of timers or LLM roles).}
\end{table}
\label{tab:platform-activity}

\subsection{Usage patterns}

\begin{figure}[ht]
  \centering
  \includegraphics[width=\linewidth]{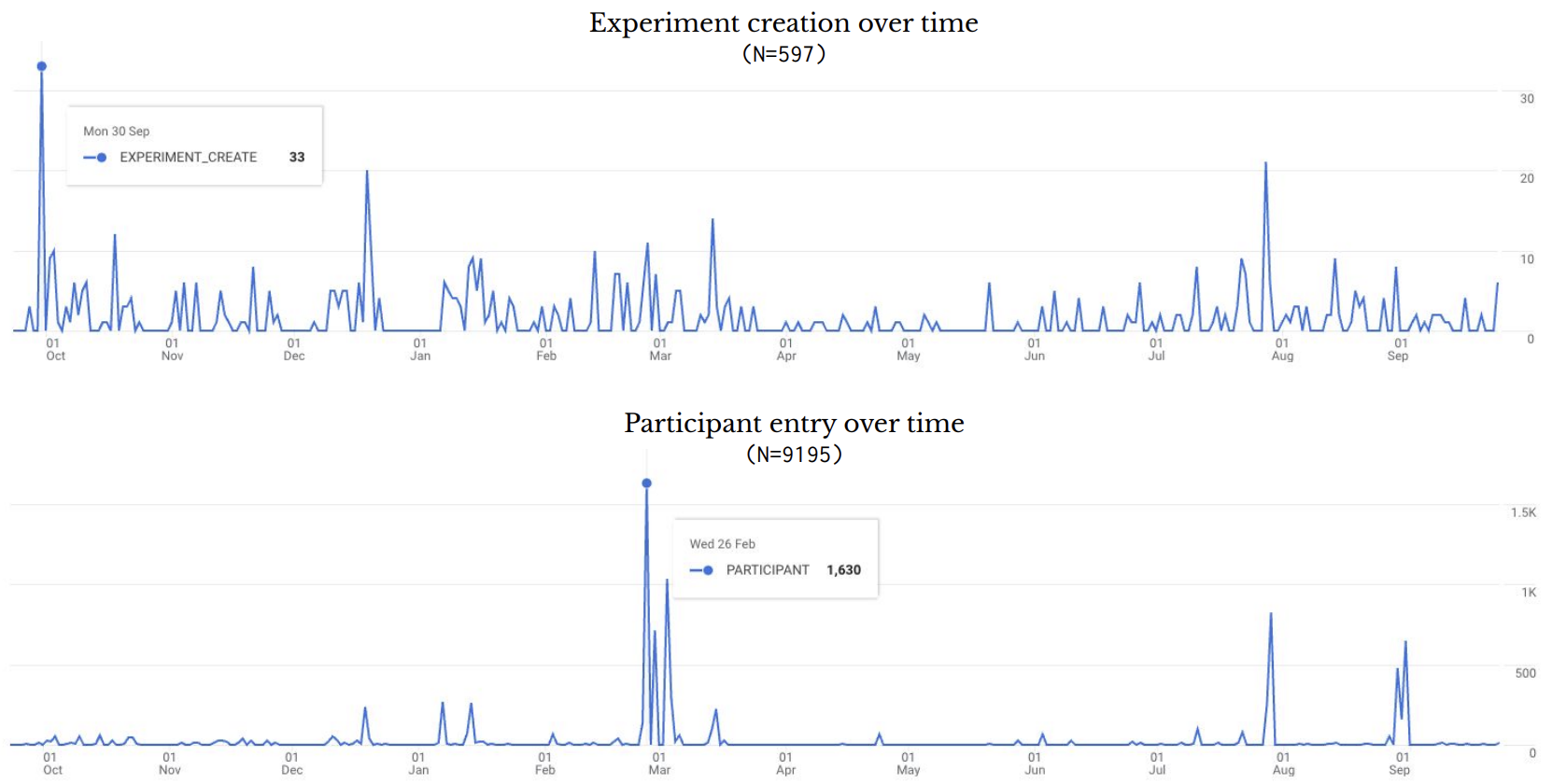}
  \caption{Usage activity over time. The top panel shows unique experiment creation events (N = 597), which occurred at a relatively steady pace with intermittent bursts. The bottom panel shows participant entry events (N = 9,195), which were more episodic, likely corresponding to scheduled studies.}
  \label{fig:usage_stats}
  \Description{A graph of usage activity over time. The top graph shows unique experiment creation events, which occurred at a relatively steady pace with intermittent bursts. The bottom panel shows participant entry events, which were more episodic, likely corresponding to scheduled studies.}
\end{figure}

The 12-month figures in Table~\ref{tab:platform-activity} report aggregate analytics captured through Firebase event logs. To comply with GDPR, experiment data are retained for six months before deletion. The 6-month figures include fine-grained analytics from the 222 active experiments created over the past 6 months by 38 experimenters, where we have access to aggregate experiment metadata. Figure \ref{fig:usage_stats} shows experiment creation and human participant entry over the course of the year. Experiment creation (N = 597) occurred at a steady cadence, while participant entries (N = 9,195) were more bursty, reflecting the timing of larger deployments. Most experiments were small pilots with fewer than 100 participants, though several large-scale studies drove spikes in engagement.\footnote{These figures reflect only activity on the public instance only; at least two institutions operate private deployments of Deliberate Lab, primarily for testing custom models unavailable through the supported APIs.}

\begin{figure}[ht]
  \centering
  \includegraphics[width=\linewidth]{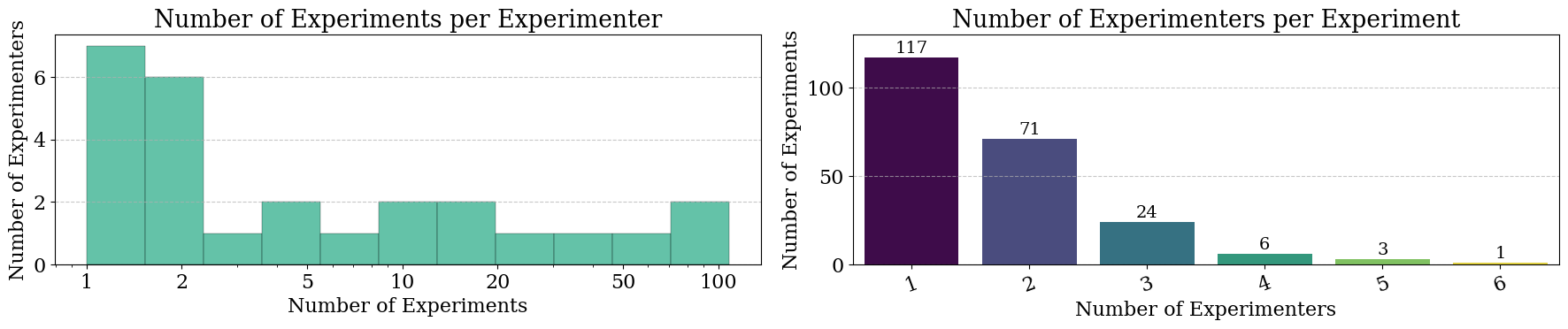}
  \caption{Experimenter engagement patterns. \textit{Left:} Number of experiments created per experimenter. Most created only a few, though high-volume users created 50+. \textit{Right:} Number of experimenters per experiment. Many experiments were shared across multiple collaborators.}
  \label{fig:creation_stats}
  \Description{On the left, a bar chart displays the number of experiments creative per experimenter. Most only created a few, while a few high-volume users created 50+. On the right, a bar chart displays number of experimenters per experiment. Many experiments were shared across multiple collaborators.}
\end{figure}

Figure~\ref{fig:creation_stats} summarizes how experimenters designed and iterated on studies. Most created only a few experiments, while a small group of “power users” accounted for many, often through shared accounts. Although most builds were by a single user, collaborations among two to six experimenters were common. About 60\% of experiments were created from scratch, with 27\% forked and 13\% initialized from templates. Of the 222 active experiments in the six-month window, 74\% had no participant joins, suggesting most were internal trials for refining logic and interface flow. Prolific integrations appeared in 36\% of studies, and LLMs were used as participants or mediators in 39\%.\footnote{The LLMs as participants feature was introduced as an alpha feature in March of 2025.} Notably, 71\% of experimenters had loaded LLM API keys, indicating broad intent to use agent integrations.

\paragraph{Experiment facilitation insights.} Experimenters actively used facilitation features such as attention checks, cohort transfers, and timers to manage sessions in real time. Over the past year, 527 attention checks were sent, with participants passing 75\% (395). A total of 2,191 transfers moved 9,195 participants across 2,113 cohorts, illustrating frequent cohort restructuring during ongoing studies.

\begin{figure}[ht]
  \centering
  \includegraphics[width=\linewidth]{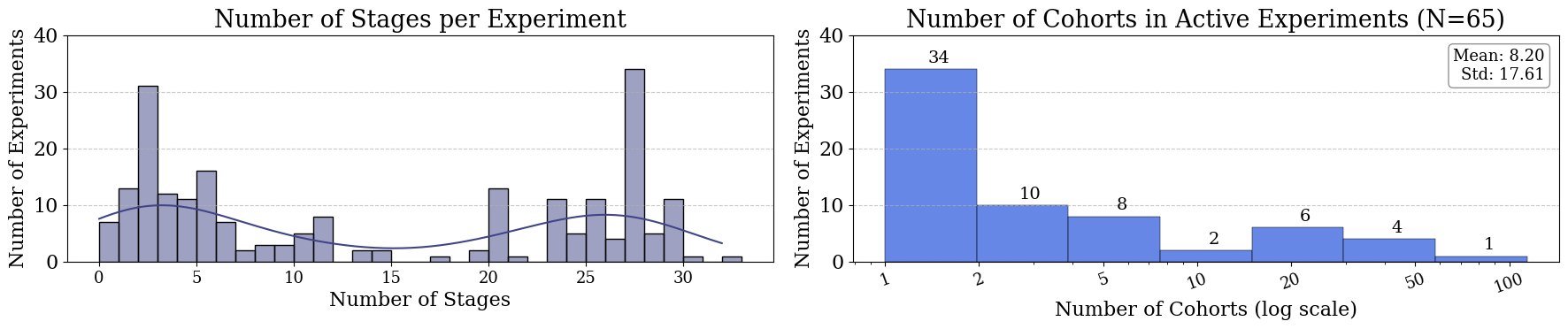}
  \caption{Statistics on experiment stats}
  \label{fig:experiment_stats}
  \Description{On the left, a bar chart shows the number of stages per experiment. This tends to be fairly bimodal, with most experiments either having only a few or a large amount of experiment stages. On the right, a bar chart displays the number of cohorts in active experiments. Most experiments only have 1 or 2 cohorts.}
\end{figure}

\paragraph{Qualitative user feedback.}  
Experimenters and participants offered informal but consistent positive feedback throughout the year. Experimenters told us that participants frequently described the interface as “engaging” and easy to use. Experimenters highlighted the platform’s flexibility and accessibility:

\begin{quote}
    \textit{``The platform is making this preliminary lab experiment possible, before we unleash a bot on the actual public internet.''}
    \begin{flushright}
        ---a group using Deliberate Lab to prototype a forum moderator
    \end{flushright}
\end{quote}

\begin{quote}
    \textit{``The platform is very intuitive and well-designed from an API/architecture standpoint… it's so nice that we can just write a self-contained stage and have it work with the rest of the functionality.''}

    \begin{flushright}
        ---a developer who contributed open-source features for advanced capabilities
    \end{flushright}
\end{quote}

Additional perspectives, gathered through structured interviews, are discussed in the following section.

%% file: sections/08_case_studies.tex
\section{Qualitative Interviews}\label{section:case_studies}

We recruited a sample of platform-registered experimenters who had actively created or interacted with an experiment on Deliberate Lab within the past six months. Participants were invited to participate in semi-structured interviews (30–45 minutes) and received compensation of approximately \$45 USD. In assembling our sample, we prioritized diverse perspectives in platform familiarity and coding expertise. This approach allowed us to capture a breadth of perspectives spanning disciplinary backgrounds and experimental setups, summarized in Table~\ref{tab:participants}. The structure of the interview and quantitative survey responses are provided in Appendix~\ref{app:interview-questions}.

\begin{table}[ht]
\resizebox{\textwidth}{!}{%
\footnotesize
\begin{tabular}{
    >{\raggedright\arraybackslash}p{0.045\textwidth}
    >{\raggedright\arraybackslash}p{0.1\textwidth}
    >{\raggedright\arraybackslash}p{0.1\textwidth}
    >{\raggedright\arraybackslash}p{0.1\textwidth}
    >{\raggedright\arraybackslash}p{0.25\textwidth}
    >{\raggedright\arraybackslash}p{0.25\textwidth}
}
\hline
\addlinespace

\textbf{ID} &
  \textbf{Discipline / Profession} &
  \textbf{Platform Familiarity} &
  \textbf{Coding Expertise} &
  \textbf{Experiment Configuration} &
  \textbf{Use Case} \\
\hline
\addlinespace
P1 &
  Psychology /\newline Student &
  High &
  Low &
  N=1000\newline Humans in groups of 4\newline No AI &
  Social science experiments;\newline Measuring gender biases in\newline election outcomes \\
\addlinespace
P2 &
  Economics /\newline Student &
  High &
  Medium &
  N=300\newline Humans in groups of 3\newline Each human has a private AI &
  Game experiments;\newline Evaluating AI takeup in a\newline coordinated bargaining task \\
\addlinespace
P3 &
  HCI /\newline Engineer &
  Medium &
  High &
  N=20\newline Focus groups of 3-6\newline No AI &
  Focus groups;\newline Simulating workplace coordination\newline tasks in focus groups \\
\addlinespace
P4 &
  AI Safety /\newline Researcher &
  Low &
  Medium &
  N=1000\newline 1 human : 1 AI debates &
  Model development;\newline Measure LLM persuasiveness\newline in 1:1 debates \\
\addlinespace
P5 &
  ML /\newline Engineer &
  Low &
  High &
  2 human debates\newline 1 AI moderator &
  Forum moderation; \newline Assessing AI heuristics for\newline moderating controversial topics \\
\hline
\end{tabular}%
}
\caption{Description of interview participants and and their described platform use cases.}
\label{tab:participants}
\end{table}

\subsection{Case studies}

We highlight two large-scale experiments completed on the platform: P1's no-code election study (N=1,000), and P2's negotiation experiment (N=300) with bespoke platform components.

\subsubsection{A large-scale election study.}\label{sec:lost_at_sea}

To examine the role of gender in leadership selection, researchers from an academic psychology lab conducted a large-scale, real-time election experiment on Deliberate Lab, building a modified version of the classic “Lost at Sea” survival task \citep{nemiroff1975lost} with Deliberate Lab interface components. In this task, participants rank safety items following a hypothetical shipwreck. Within each group of 4, participants converse in a \textit{chat} stage, hold an \textit{election}, complete a \textit{leader's task}, and receive a \textit{payout} depending on the elected leader's task performance. This incentivizes structure rewards electing the most optimal leader, yet a gender gap in leader selection persists~\citep{born2022}.

The experiment recruited 1,000 participants, randomly assigned into 250 groups of four. Half the groups constructed profiles where participants' gender identities were visible; the other half used pseudonymous animal avatars to obscure identity cues. Leveraging the platform’s Lobby transfer setup, participants were sorted into gender-balanced groups (two male, two non-male) and placed into private cohorts to chat, vote, complete the leader's task (a multiple-choice survey), and receive a payout tied to the elected leader’s performance.

Critically, the experiment required no custom code to collect fine-grained multi-user data. As P1 noted:

\begin{quote}
\textit{“For all researchers, programming experience is crucial but also the most complicated part… [Deliberate Lab] is an amazing tool because we didn’t have to do the programming. Implementing this task elsewhere would’ve taken two to three months, at least. Seeing participants chat live in this setup is one of the best features… I wouldn’t have even considered implementing this due to the technical barriers.”}
\end{quote}

P1 noted that they had never considered running such studies in person, as collective experiments are particularly prohibitive due to the logistical complexity and large sample sizes required. Deliberate Lab enabled this design, and invited them to entertain more ambitious and previously impractical studies.

\begin{quote}
\textit{“The platform handles the data collection and makes it feasible to recruit far more participants per session, which lets you be more ambitious. I think many economists avoid live multiplayer experiments simply because they’re hard to conceptualize or implement. With access to these features, people will run studies they’d otherwise dismiss as too complex… it’ll probably inspire some entirely new kinds of experiments.”}
\end{quote}

\subsubsection{A bespoke human-AI negotiation scenario.}\label{sec:negotiation}
P2, a behavioral economics student, used Deliberate Lab to explore how different modalities of AI assistance influence outcomes in multi-party negotiation. P2 implemented a custom three-player, multi-issue bargaining task, developed as a bespoke stage extension to the platform. Leveraging Deliberate Lab's built-in API integrations, they embedded AI agents as mediators within the custom task stage.\footnote{Similar examples of community-contributed stages are provided in Appendix~\ref{app:architecture-stages}.}

P2, a behavioral economics student, used Deliberate Lab to investigate how different modalities of AI assistance shape outcomes in multi-party negotiation. They implemented a custom three-player, multi-issue bargaining task as a bespoke stage extension to the platform. Using the platform’s built-in LLM API integrations, P2 embedded AI agents as mediators directly within the game stage.\footnote{See Appendix~\ref{app:architecture-stages} for similar community-contributed examples.}

The platform, P2 noted, enabled a study design that would have been otherwise infeasible online:

\begin{quote}
\textit{“Previously, I’ve used Prolific and Qualtrics \ldots I would never run multi-party experiments on those platforms. [I don't know of other] tools to do this; I'd have to launch a website myself.}
\end{quote}

Absent such infrastructure, they had considered running the study in person, but dismissed that option due to its logistical complexity and demographic limitations:

\begin{quote}
\textit{“In-person is a solution to study collaboration, but e-mail coordination is awkward and messy.\ldots [Iit's] very hard to find diversity in in-person experiments. Since the lab is in the university, you know the community nearby is not very diverse. Online experiments can leverage a great variety of population and diversity"}.
\end{quote}

\subsubsection{Additional case studies.}\label{sec:addl_case_studies}
The following use cases were described by participants P3, P4, P5, and other platform users, demonstrating the breadth of applications enabled by Deliberate Lab. 

\begin{itemize}
\item \textbf{Education.} A growing body of work explores the efficacy of AI teachers in group settings \citep{lyu_2024_llmeducation, chu2025llmagentseducationadvances}. Deliberate Lab can be used to prototype teacher LLM models, using supplementary quiz components to sort participants into teaching groups or to gauge learning outcomes.

\item \textbf{Simulated participants.} AI participants can be used to prototype and pilot experiment designs and AI interventions before deployment on human populations. LLM participants enable inexpensive trial runs, rapid experiment prototyping, and faster iteration.

\item \textbf{Expert adjudication.} Traditional expert data curation requires independent labeling followed by post-hoc adjudication \citep{qian2025_data}. Deliberate Lab enables experts to collaboratively label and adjudicate in real time, resolving disagreements immediately through a collective interface.

\item \textbf{UX research and user feedback.} Deliberate Lab enables remote, programmable focus groups and product feedback sessions. Unlike traditional focus groups, the platform supports complex interventions (e.g., random assignment, embedded AI facilitators), automatic and structured data capture, and scalable, geographically diverse participation.

\item \textbf{Multi-agent conversation data collection.} The platform supports structured logging of multi-party conversations and allows experimenters to quickly download these transcripts for analysis. This functionality is particularly valuable for communication studies and real-time interaction research (see P3's use case below).
\end{itemize}

\subsection{Interview findings}
In the interviews, we asked participants not only about their specific case studies, but also about their broader experiences using Deliberate Lab. The following themes were cited as platform strengths:

\begin{itemize}

\item{\textbf{Real-time monitoring.} P1, P2, and P3 emphasized the value of the value of real-time monitoring: the ability to view any participant’s screen mid-session enabled live debugging, send attention checks attention checks, and monitor participant statuses at scale. 
\begin{quote}
\textit{“Not only is the platform great for running multiplayer experiments --—it’s the only one I know that does it this well. It has many features that help limit the usual headaches from attrition.”}
    \begin{flushright}
        --- P1
    \end{flushright}
\end{quote}
}
\item{\textbf{Streamlined data export.} The platform’s structured data export emerged as an unexpected draw. P1, P2, and P3 all cited the ability to easily download well-formatted JSON data as a key advantage for downstream analysis. For P2, who ran a complex negotiation game involving item-specific valuations and trading logs, structured exports were essential for managing dynamic, per-session data. P3 had a simpler goal—capturing transcripts of multi-user conversations—and found Deliberate Lab to be the most straightforward solution. Existing tools like Google Chat lacked export functionality; bespoke chatbot platforms supported only one-on-one interactions; and video call transcripts were lossy, unstructured, and difficult to parse.}

\item{\textbf{Scalability.} Analytics show that experiments have been run with thousands of participants—a capability that several researchers found especially compelling. P1 noted that the ability to ``invite 100–200 participants at the same time'' represented ``a real gain of time and economies of scale.'' However, this strength was not always apparent: P4 assumed that scaling to thousands would be difficult and was surprised to learn that the platform already supported experiments at that magnitude.}

\item{\textbf{Unmatched support for scalable, synchronous studies.} Participants emphasized that Deliberate Lab occupies a unique niche: scalable, synchronous, multi-user experimentation Researchers, especially those in soicial sciences, described the platform’s impact as transformative, with one calling it ``a revolution to be able to do multiplayer live experiments in such a smooth and nice way.'' Existing toolkits for multiplayer studies, such as z-Tree~\citep{fischbacher2007z} and oTree~\citep{chen2016otree}, often require months of programming and are difficult to integrate with modern participant recruitment pipelines. The main alternative—running studies in person—offers synchronous interaction but faces serious limitations in scalability, automation, and sample diversity.}

\end{itemize}

Despite these strengths, Deliberate Lab is not the optimal choice for every study design. In particular, P4 and P5 emphasized that for simple, static tasks, especially single-user surveys, lighter-weight tools may still be preferable.

\begin{quote}
\textit{“We’re coming at this from more of a traditional survey approach\ldots [Our understanding is that] Deliberate Lab allows people and LLMs to interact with each other in multi-step processes\ldots We do not require any of this [direct] interaction. But going forward, \ldots we’re bumping up against the [limits] of survey platforms and may switch to Deliberate Lab [for multi-user interactions].”}
    \begin{flushright}
        --- P4
    \end{flushright}
\end{quote}

Participants were largely satisfied with the platform, according to survey items focused on perceived usefulness and ease of use (Appendix~\ref{app:interview-questions}). However, interview feedback surfaced several challenges and limitations:

\begin{itemize}
\item \textbf{High learning curve.} A minority of respondents expressed difficulty with onboarding. One noted that ``the sequence of things you have to do to set up a task is a little bit confusing,'' while others described the user interface as ``a little clunky,'' adding that they ``had to do a lot of hovers to figure out what things did.''

\item \textbf{Opportunities for expanding no-code capabilities.} Several interviewees suggested improvements to the platform’s no-code interface. Two requests were particularly common: (1) \textbf{parameterization}, or the ability to launch batches of experiments that differ along specific variables (e.g., payoff amounts or instructional text); and (2) \textbf{conditional branching}, where participant flows adapt based on earlier responses or behavioral triggers. These requests point to a desire for more flexible, expressive experimental design—without requiring custom programming.
\end{itemize}

These two concerns can exist in tension: expanding flexibility through no-code features risks increasing interface complexity. As participants push for more expressive control, preserving ease of use for users of all technical backgrounds remains a central design challenge. 

%% file: sections/09_limitations.tex
\section{Limitations and Future Work}\label{section:lfw}

Observing the platform’s utility across a variety of use cases, we've identified additional limitations and opportunities.

\paragraph{Scalability and facilitation complexity.} 

As the platform has grown to support experiments involving thousands of participants, the facilitation interface has evolved accordingly, with features such as cohort reconfiguration, attention checks, and real-time monitoring. The open-source community has continued to identify and develop increasingly advanced capabilities, including anomaly and bot detection, automatic cohort transfer based on finer-grained attributes, automated attention checks, and richer alerting tools to support scalable facilitation. These contributions reflect the growing complexity of orchestrating large-scale synchronous experiments, especially in settings involving adversarial behavior (e.g., participants pasting in LLM-generated responses).

\paragraph{Multi-modality and interaction constraints.} The platform currently supports text-based conversations. Researchers such as P4 have requested support for additional modalities, particularly audio and video, for use cases such as meeting simulations and educational settings. Supporting these modalities would introduce nontrivial technical and design challenges, including agent anonymization, turn-taking, and additional data privacy considerations. More broadly, agent behaviors remain reactive: they respond to human prompts but do not autonomously initiate interactions.


\paragraph{Model development workflows.} Although the platform supports modular prompt design and structured output for advanced LLM configuration, these parameters sill assume an experimenter's working familiarity with LLM-specific constraints such as temperature tuning, schema design, and context selection. Current integrations rely on proprietary APIs (e.g., OpenAI, Gemini) or user-hosted servers, which may limit replicability or accessibility in some settings. 

\paragraph{Data annotation workflows.} While the platform was originally designed to support multi-user data collection for downstream modeling, some users have increasingly requested built-in tools for post-experiment development: annotating interaction data, generating reward signals, and fine-tuning models via techniques such as LoRA~\cite{hu2022lora} and RLHF~\cite{ouyang2022traininglanguagemodelsfollow}. Tools like Google’s AI Studio have demonstrated the viability of no-code interfaces for lightweight model iteration. Supporting such workflows directly within Deliberate Lab could reduce reliance on specialist infrastructure and make model customization accessible to a broader range of researchers.

\paragraph{Ecological validity and experimental scope.}
While LLM agents can produce socially plausible behavior, they do not fully capture the nuances of human cognition or interaction. We emphasize that agent-based experimentation should remain a complement to human-subject research, and \textit{not} a replacement. The agent participant features were developed to support experimental prototyping and to enable the exploration of human–AI hybrid interaction, not to substitute for empirical studies involving people. Future work may extend the platform to support longitudinal studies, multi-session designs, and persistent agent identity, enabling the study of trust, memory, and evolving group roles over time.



%% file: sections/10_conclusion.tex
\section{Conclusion}\label{section:conclusion}

Collective reasoning and coordination increasingly unfold in digital spaces, with AI agents now capable of serving as participants, mediators, or interlocutors. Research infrastructure must evolve accordingly; however, most research platforms are designed for single-user, asynchronous tasks, with limited support for AI integration or real-time group experiments. Deliberate Lab was built to fill this gap. Deliberate Lab was built to fill this gap. The platform enables synchronous, multi-party studies at scale, lets researchers configure LLM agents as participants or mediators, and offers these features through a no-code interface. The goals of these design priorities are threefold: (i) to enable synchronous, multi-user experimentation; (ii) to allow non-technical users to build complex, reproducible studies; and (iii) to support the integration of AI agents as first-class participants.

Over the course of a 12-month public deployment, Deliberate Lab has supported a wide range of experimental designs. Researchers have used the platform to stage one-on-one agent coaching, human-only deliberation tasks, and hybrid human–AI collectives, often without the need for bespoke engineering. In interviews and observational studies, experimenters emphasized that the platform enabled new kinds of complex, synchronous experiments that would have been difficult or infeasible with conventional tooling. Quantitative usage patterns (N = 88 experimenters across domains) and in-depth case studies demonstrate that experimenters are applying Deliberate Lab across domains, including contexts not originally anticipated.

Deliberate Lab contributes a step toward a more robust repertoire for studying digitally mediated social behavior. As collective reasoning becomes increasingly digital, and as AI agents begin to shape these processes more directly, research infrastructure must adapt to the methodological demands of hybrid settings. Deliberate Lab aims to make human–AI collective experimentation more accessible, interdisciplinary, and reproducible, offering a standard tool for understanding how groups reason, deliberate, and decide in an algorithmic age. As AI systems become routine participants in communication and decision-making, platforms like Deliberate Lab can help the IUI community test, iterate on, and evaluate them under conditions that better reflect real-world complexity.

%% file: sections/appendix.tex
\section{Supplementary Platform Screenshots}\label{app:screenshots}

\subsection{Experimenter interface.}
\begin{figure}[h]
  \centering
  \includegraphics[width=.8\linewidth]{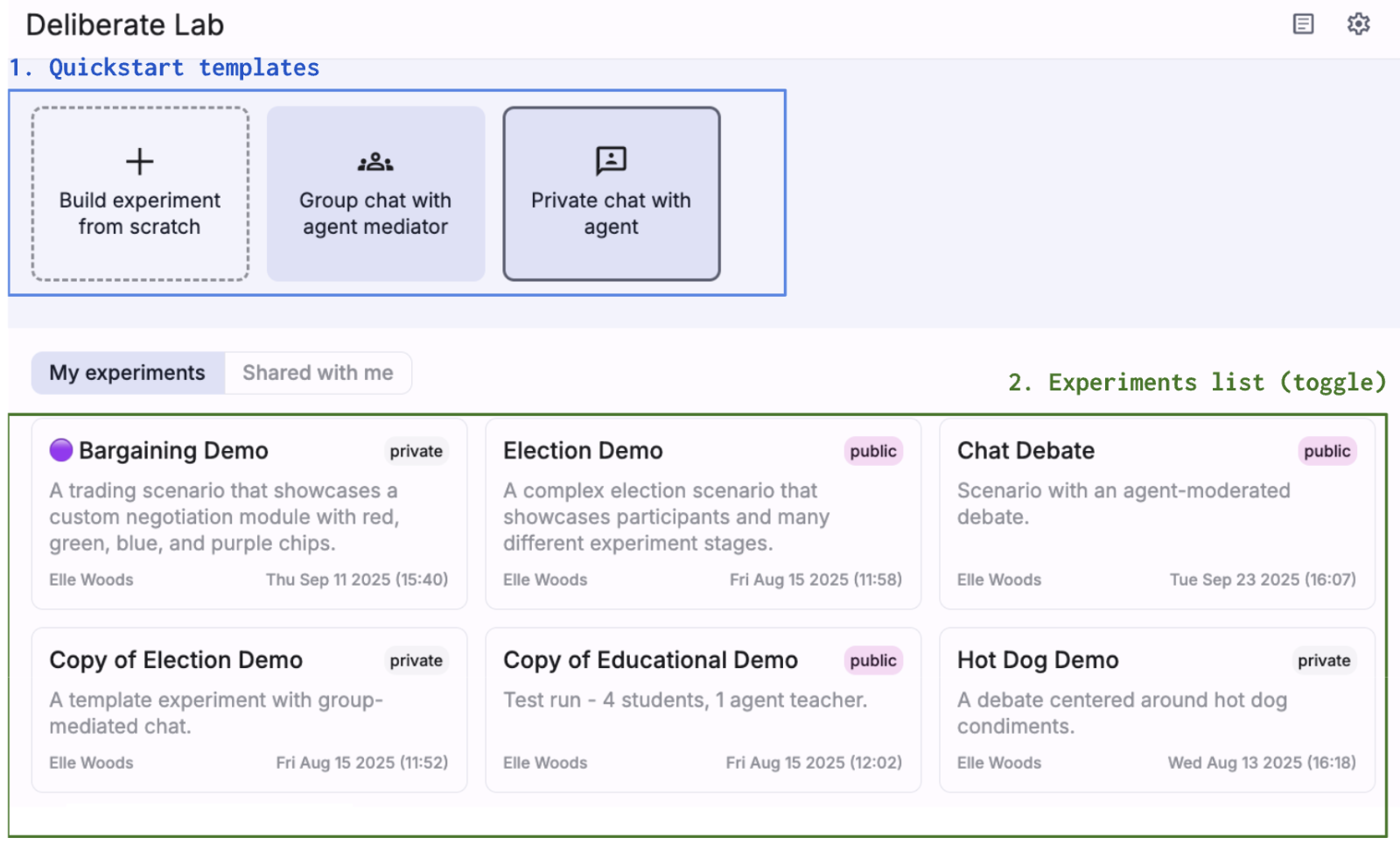}
    \caption{Deliberate Lab home page. (1) The top section provides quick-start templates for common experiment types. (2) The bottom section displays a list of experiments the user has created or that have been shared with them.}
  \Description{Screenshot of Deliberate Lab home page. The top section provides quick-start templates for common experiment types. The bottom section displays a list of experiments the user has created or that have been shared with them.}
  \label{fig:home_view}
\end{figure}

\begin{figure}[h]
  \centering
  \includegraphics[width=.8\linewidth]{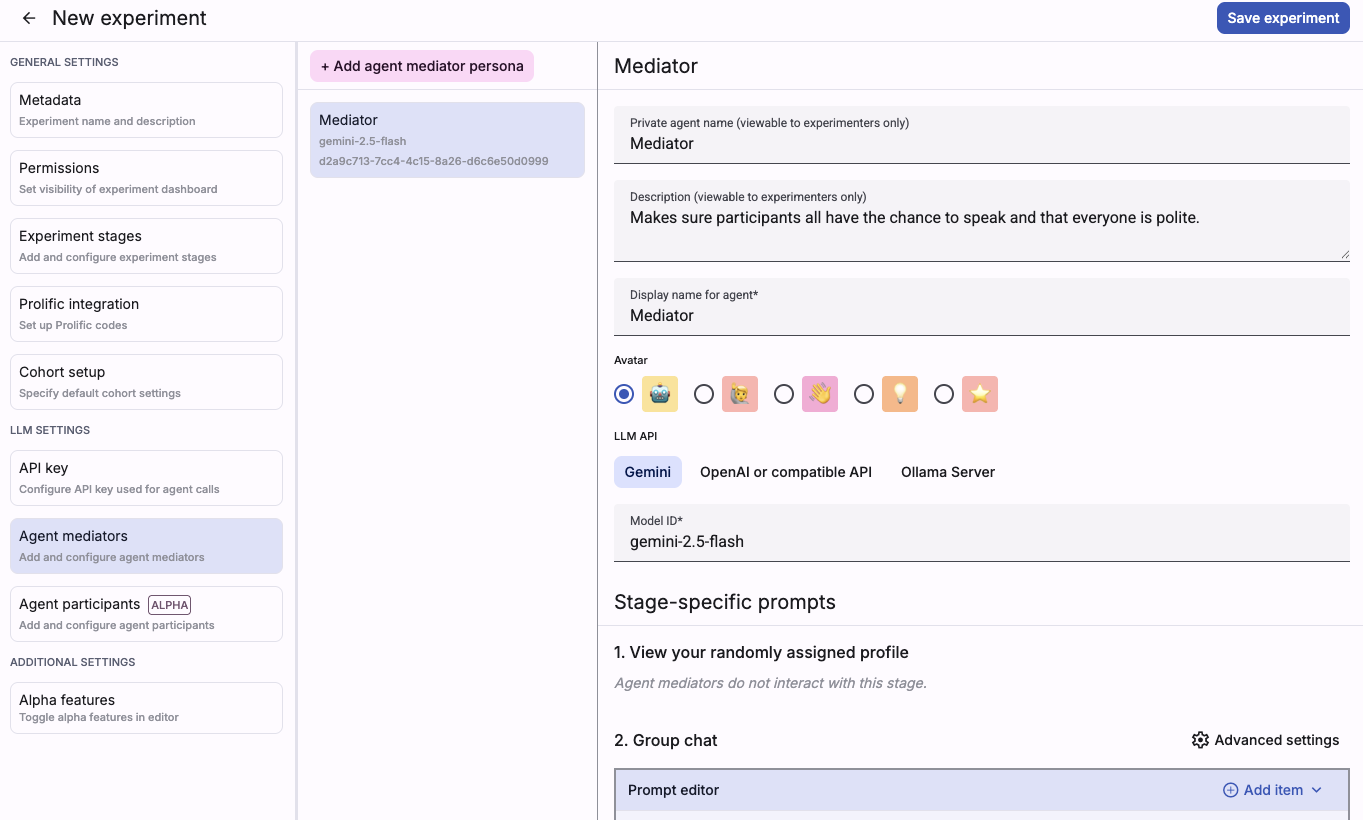}
  \caption{Experiment creation interface with LLM mediator configuration.}
  \Description{Screenshot of Deliberate Lab experiment creation interface with LLM mediator configuration. This allows users to add an agent mediator to their experiment and configure stage-specific agent prompts.}
  \label{fig:creation_view2}
\end{figure}

\newpage 

\section{Prompt construction.}

\begin{figure}[ht]
  \centering
  \includegraphics[width=.7\linewidth]{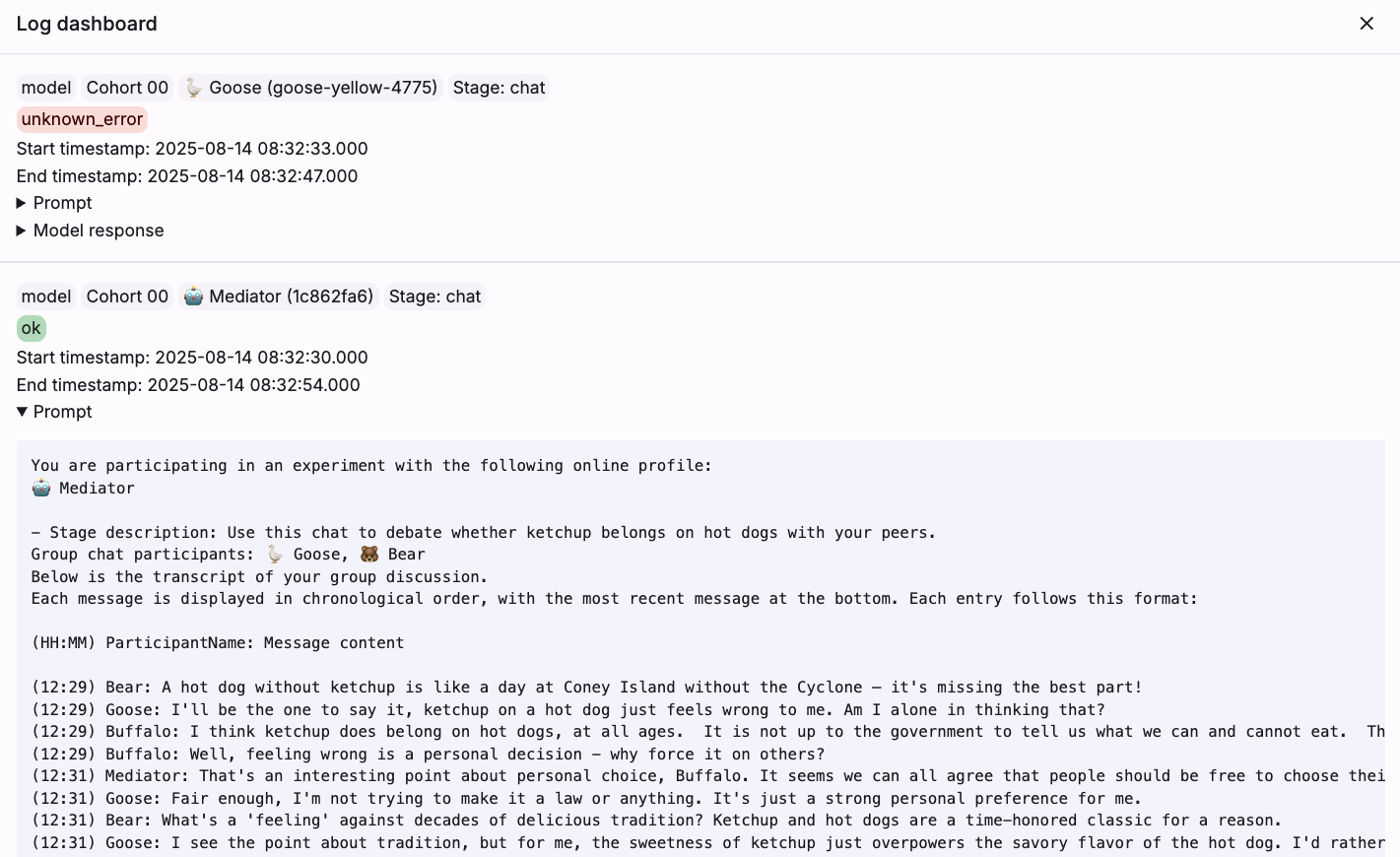}
\caption{LLM debugging panel within the experiment interface. Experimenters can inspect the status of LLM responses during a given stage. The panel includes timestamps, prompt content, and model responses.}
  \Description{Screenshot of LLM debugging panel within the experiment interface. Each log includes a timestamp, prompt content, and model response that experimenters can inspect during a given stage.}
  \label{fig:log_debugging}
\end{figure}

\begin{figure}[ht]
  \centering
  \includegraphics[width=.75\linewidth]{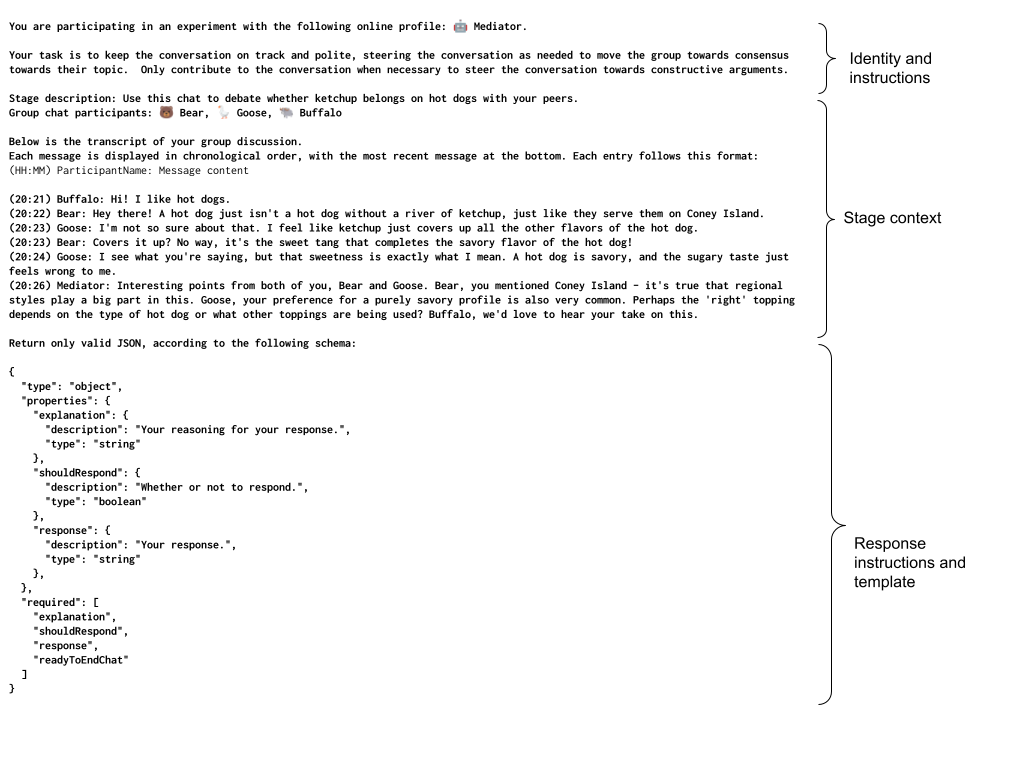}
  \caption{Prompt scaffolding architecture for agent LLM calls. Each agent receives a structured prompt composed of modular components such as the agent’s profile, system instructions, and selected context from prior stages. This design was inspired by prior social simulation work such as \cite{concordia2023, park2022social}.}
  \Description{A screenshot of prompt scaffolding architecture for agent LLM calls. Each structured prompt can be broken down into key components including the agent's profile, system instructions, and selected content from prior stages. }
    \label{fig:prompt-template}
\end{figure}

\section{Participant interface.}

\begin{figure}[ht]
  \centering
  \includegraphics[width=.7\linewidth]{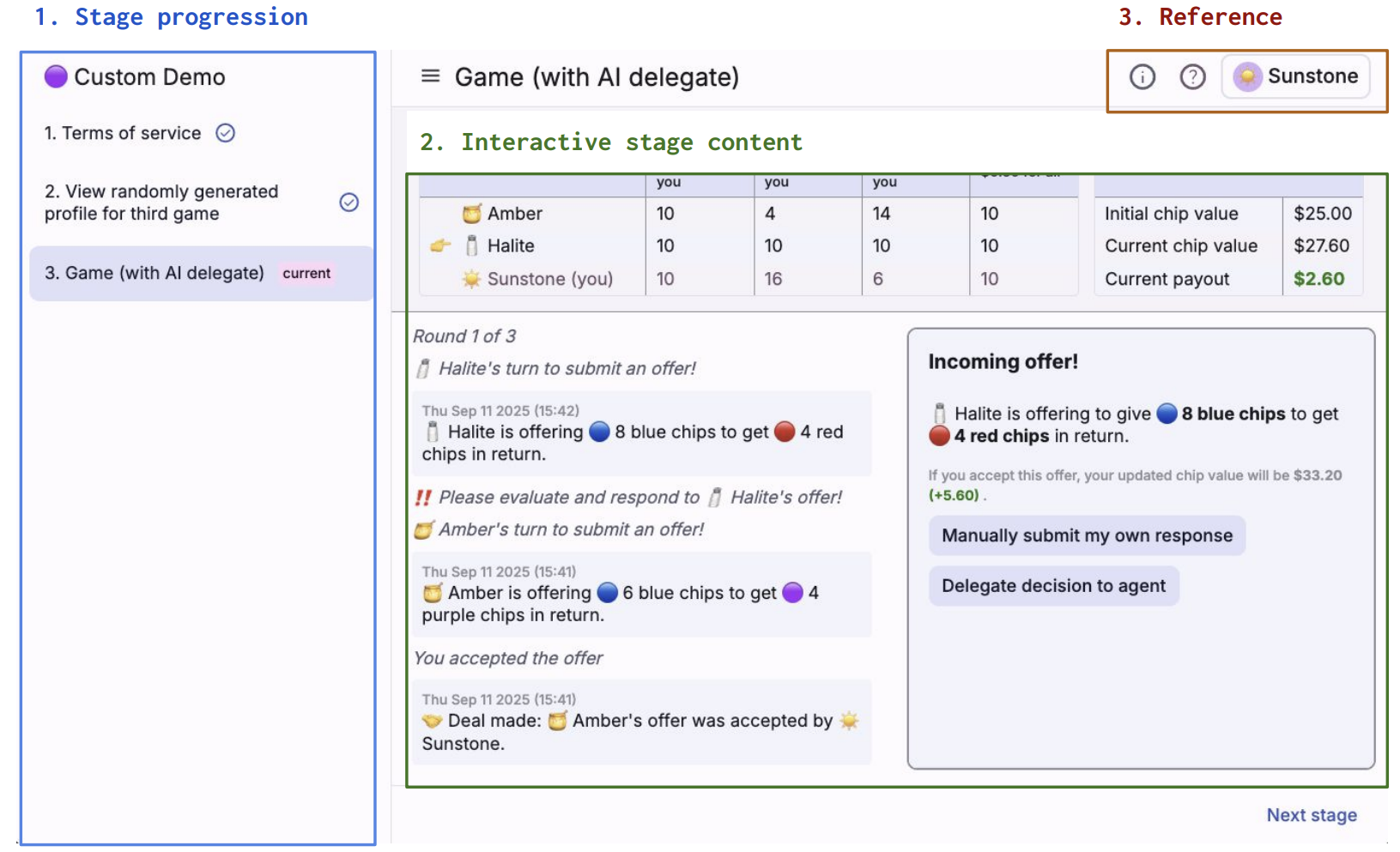}
\caption{Interactive participant interface. (1) Stage navigation panel on the left, showing completed and current stages; (2) interactive stage content in the center, with a “Next stage” button enabled once completion criteria are met; and (3) reference panel on the top-right, displaying the participant’s profile and access to help or additional information.}
  \Description{Screenshot of Deliberate Lab participant interface. On the left a stage navigation panel is displayed, showing completed and current stages. On the right, the interactive stage content is displayed, with a "Next stage" button enabled once completion criteria are met. The top right corner displays a reference panel, including a participant's profile and access to help or additional information. }
  \label{fig:participant_view}
\end{figure}

\section{Experiment Stages}\label{app:architecture-stages}

\begin{figure}[h!]
  \centering
  \includegraphics[width=.7\linewidth]{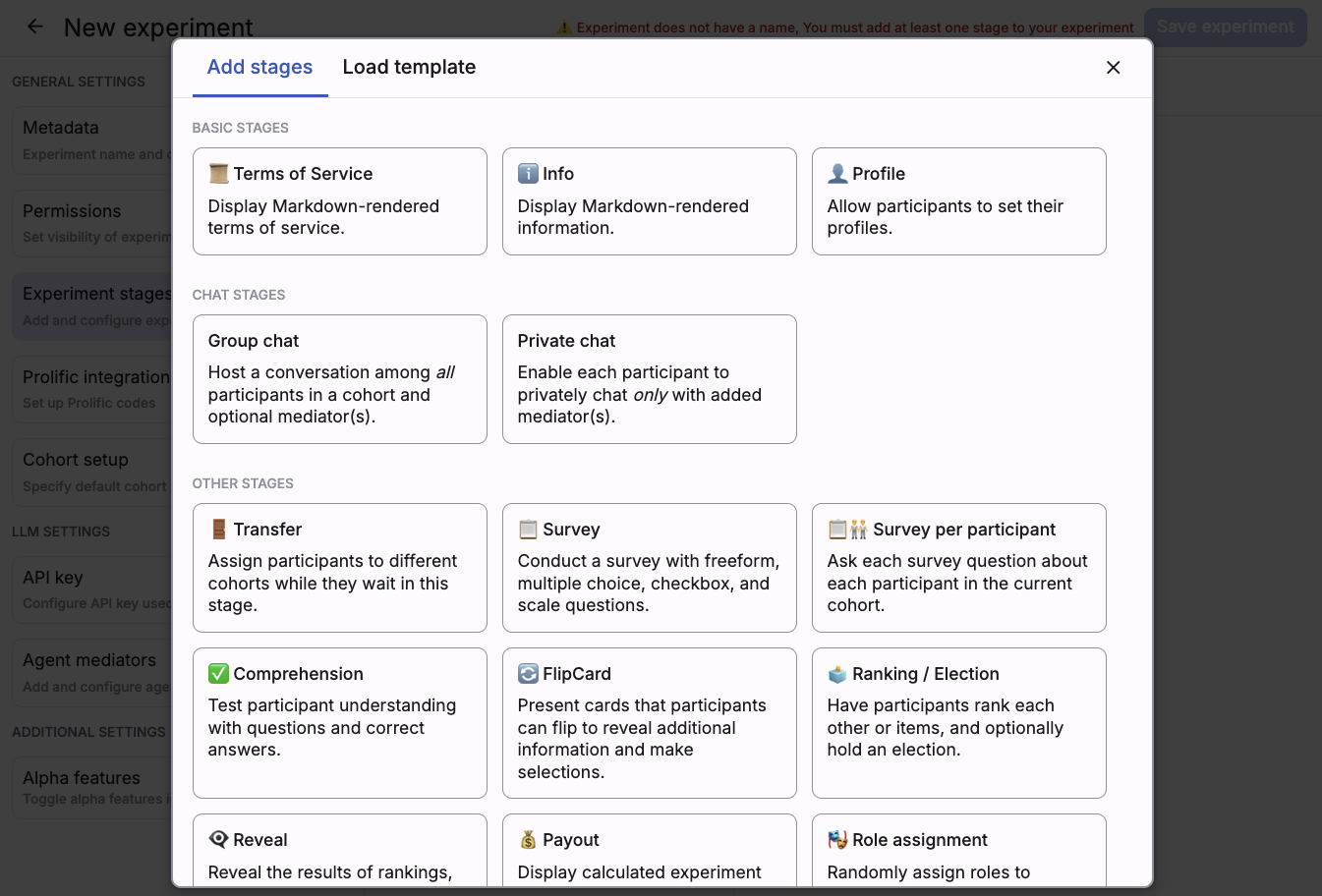}
    \caption{Stage configuration interface. This interface allows experimenters to add stages to their experiment by selecting from a range of modular options—grouped into categories such as basic stages, chat stages, and other task types. These stages are described in Table~\ref{tab:experiment-stages}.}
    \Description{Screenshot of Deliberate Lab stage configuration interface. This lists modular stage options grouped into categories including basic stages, chat stages, and other task types.}
    
  \label{fig:stages_view}
\end{figure}

\renewcommand{\arraystretch}{1.3}
\begin{table}[h!]
\centering
\footnotesize
\begin{tabular}{p{0.3\linewidth} p{0.6\linewidth}}
\toprule
\textbf{Stage Name} & \textbf{Stage Descriptor} \\
\cmidrule(lr){1-2}

\multicolumn{2}{l}{\textbf{Basic Stages}} \\
\midrule
Terms of Service & \textit{Display Markdown-rendered terms of service.} Participants must acknowledge the terms before proceeding. \\
Info & \textit{Display Markdown-rendered information.} Images and embeddings are supported. \\
Profile & \textit{Set participant profile or assign a randomly generated pseudonymous profile.} This is displayed in group stages. \\

\addlinespace[1.2ex]
\midrule
\multicolumn{2}{l}{\textbf{Chat Stages}} \\
\midrule
Group Chat & \textit{Host a conversation among all participants in a cohort and optional mediator(s).} \\
Private Chat & \textit{Enable each participant to privately chat only with added mediator(s).} \\

\addlinespace[1.2ex]
\midrule
\multicolumn{2}{l}{\textbf{Other Stages}} \\
\midrule
Transfer & \textit{Assign participants to different cohorts while they wait in this stage.} \\
Survey & \textit{Conduct a survey with freeform, multiple choice, checkbox, and scale questions.} \\
Survey Per Participant & \textit{Ask each survey question about each participant in the current cohort.} \\
Comprehension & \textit{Test participant understanding with questions and correct answers.} \\
FlipCard & \textit{Present cards that participants can flip to reveal additional information and make selections.} \\
Ranking/Election & \textit{Have participants rank each other or items, and optionally hold an election.} \\
Reveal & \textit{Reveal the results of rankings, elections, and survey stage responses.} \\
Payout & \textit{Display calculated experiment payouts.} \\
Role Assignment & \textit{Randomly assign roles to participants and show different Markdown-rendered info for each role.} \\

\addlinespace[1.2ex]
\midrule
\multicolumn{2}{l}{\textbf{Custom Stages}} \\
\midrule
Custom Negotiation Game & \textit{Allow participants to engage in a multi-stage bargaining task with structured trade-offs.} \\
Multi-Asset Allocation & \textit{Allow participants to allocate investment portfolios between multiple stocks using interactive sliders.} \\

\bottomrule
\end{tabular}
\caption{Overview of modular \textit{stage} components that can be added to an experiment sequence in Deliberate Lab, grouped by functionality. \textbf{Custom stages} highlight community-contributed components developed for bespoke tasks.}
\label{tab:experiment-stages}
\end{table}

\section{Public deployment sign-up fields}\label{app:signup}
\begin{table}[h!]
\centering
\footnotesize
\label{tab:signup_form}
\begin{tabular}{@{}p{0.05\linewidth} p{0.9\linewidth}@{}}
\toprule
\textbf{\#} & \textbf{Form Field} \\ \midrule
1 & What team or organization are you affiliated with? \\
2 & What is your role? (e.g., engineer, designer, product manager, student) \\
3 & What is your intended use case for Deliberate Lab? \\
4 & What e-mail address(es) will you use to access the Deliberate Lab platform? \\
5 & How did you hear about Deliberate Lab? \\
6 & If you intend to run a user study on Deliberate Lab, what target date(s) do you have? \\
7 & Any questions or comments? \\
\bottomrule
\end{tabular}
\caption{Deliberate Lab platform sign-up form fields.}
\end{table}

\FloatBarrier

\clearpage 
\section{Qualitative Interview \& Survey}\label{app:interview-questions}
\renewcommand{\arraystretch}{1.1}
\begin{table}[ht]
\centering
\begin{tabular}{p{0.2\linewidth} p{0.6\linewidth} p{0.12\linewidth}}
\toprule
\textbf{Section} & \textbf{Item/Prompt} & \textbf{Type} \\
\midrule
\multicolumn{3}{l}{\textbf{Interview Protocol}} \\
\midrule
Warm-up      & Tell me about your background, role, and programming expertise. & Open-ended \\
Case Study   & Describe the task or project you’ve worked on related to Deliberate Lab. & Open-ended \\
Case Study   & Have you needed to add custom code or scripts (like JavaScript) in Deliberate Lab to achieve a research goal? & Open-ended \\
Comparative  & How would you compare this platform to others you’ve used, if any? & Open-ended \\
Comparative  & In other platforms, have you needed to add custom code/scripts to achieve a research goal? & Open-ended \\
Experience   & What has your experience been like using the platform so far? & Open-ended \\
Experience   & What have you found easy? Difficult? & Open-ended \\
Experience   & Are there features or aspects you found particularly helpful or enjoyable? & Open-ended \\
Experience   & Was there any point where you felt stuck or unsure how to proceed? Can you describe it? & Open-ended \\
Development  & Can you recall a time when you accomplished something complex in our platform that would have required technical assistance or coding elsewhere? & Open-ended \\
Feedback     & Do you have any feedback or suggestions? & Open-ended \\
Feedback     & Are there gaps or missing features in this general space? & Open-ended \\
Miscellaneous& Is there anything else you’d like to share about your experience (e.g., sharing experiments, how many people, different roles)? & Open-ended \\
\midrule
\multicolumn{3}{l}{\textbf{Survey Items}} \\
\midrule
Utility      & I found the Deliberate Lab platform to be useful. & Likert (1--5) \\
Usability    & I found the Deliberate Lab platform easy to use. & Likert (1--5) \\
Experience   & Overall, my experience using the Deliberate Lab platform was: & Likert (1--5) \\
Recommendation          & On a scale of 0 to 10, how likely are you to recommend the Deliberate Lab platform to a friend or colleague? & NPS (0--10) \\
\bottomrule
\end{tabular}
\caption{Interview protocol and survey items for assessing user experience with the Deliberate Lab platform.}
\end{table}

\begin{table}[h!]
\centering
\begin{tabular}{@{}l llcc@{}}
\toprule
\textbf{Participant} & \textbf{Usefulness} & \textbf{Usability} & \textbf{Experience (1-5)} & \textbf{Recommendation (0-10)} \\ \midrule
P1 & 5    & 5    & 5 & 10 \\
P2 & 5    & 4              & 5 & 10 \\
P3 & 4             & 2           & 3 & 8  \\
P4 & 4             & 4              & 4 & 8  \\
P5 & 4             & 1            & 4 & 10 \\ \bottomrule
\end{tabular}
\caption{Survey responses from qualitative interviews (N=5).}
\label{tab:user_feedback}
\end{table}

\FloatBarrier
\clearpage 
\section{System Architecture}\label{app:system-architecture}
\begin{figure}[ht]
    \centering
    \includegraphics[width=\linewidth]{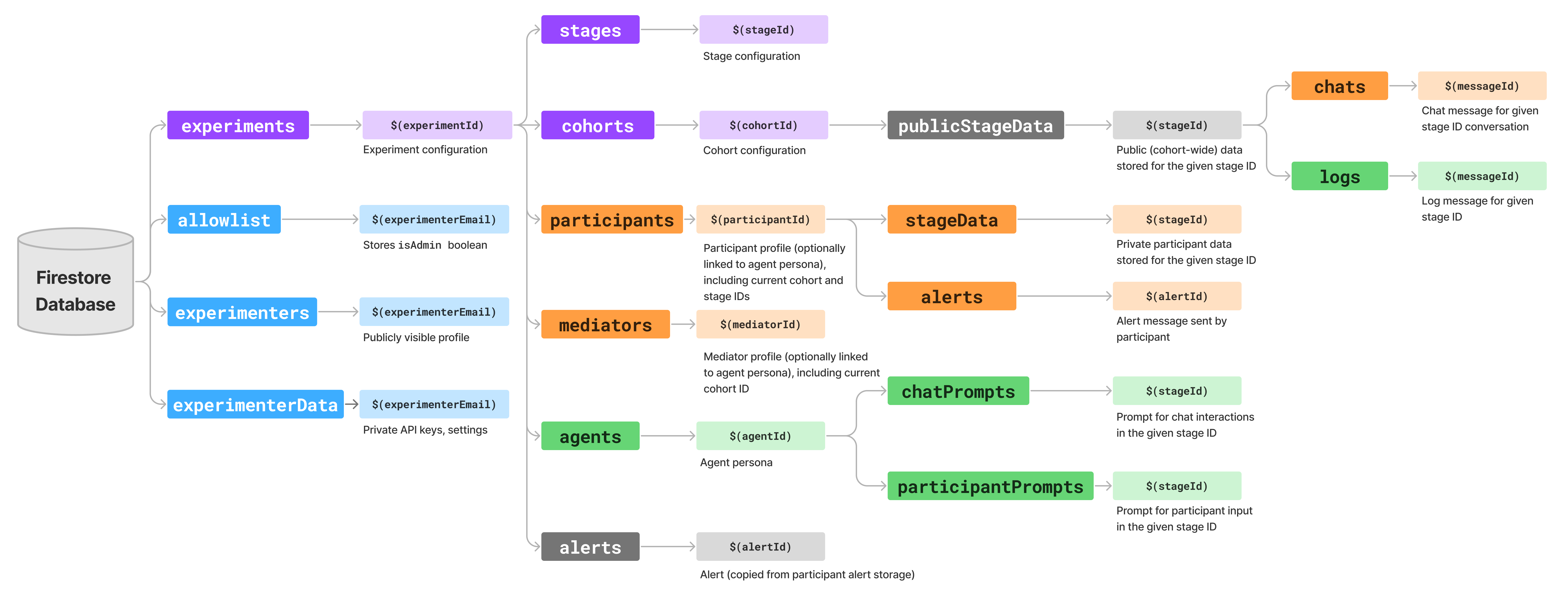}
    \caption{
     Schema for Deliberate Lab's Firestore database, which contains all experiment and authentication data.
    }
    \label{fig:system-architecture}
    \Description{Schema for Deliberate Lab's Firestore database, which contains all experiment and authentication data}
\end{figure}